\documentclass[11pt,epsf]{article}
 \usepackage{amsmath}
 \usepackage{graphicx}
 \graphicspath{{fig}}

 \usepackage[merge,numbers,compress]{natbib}
 \usepackage[T1]{fontenc}
 \usepackage{booktabs}
 \usepackage{xcolor}
 \usepackage{xspace}
 \usepackage{subfigure}
 \usepackage{dcolumn}
 \usepackage{placeins}
 \usepackage{amsfonts}
 \usepackage{yhmath}
 \usepackage[colorlinks=true,citecolor=blue!50!black,linkcolor=black]{hyperref}
 \usepackage{caption}
\usepackage[subrefformat=parens,position=top,skip=-15pt,margin=15pt,justification=justified,singlelinecheck=false]
 {subcaption}
\usepackage[normalem]{ulem}
 
\usepackage{siunitx}
\usepackage{geometry}
\usepackage{pdflscape}
\usepackage{ifthen}
\usepackage[capitalize]{cleveref}

\usepackage{comment}

\usepackage{slashed}
\usepackage{amsmath}
\usepackage{amsfonts}
\usepackage{amssymb}

\usepackage{tabularray}
\UseTblrLibrary{amsmath}

\def\citere#1{\mbox{Ref.~\cite{#1}}}

\newcommand{\newc}{\newcommand}
\newc{\beq}{\begin{equation}}
\newc{\eeq}{\end{equation}}
\newc{\beqn}{\begin{eqnarray}}
\newc{\eeqn}{\end{eqnarray}}
\newc{\bit}{\begin{itemize}}
\newc{\eit}{\end{itemize}}
\newc{\ben}{\begin{enumerate}}
\newc{\een}{\end{enumerate}}
\newc{\bce}{\begin{center}}
\newc{\ece}{\end{center}}
\newc{\bfi}{\begin{figure}}
\newc{\efi}{\end{figure}}

\newcommand{\Pe}{\ensuremath{\text{e}}\xspace}

\newcolumntype{.}{D{.}{.}{-1}}
\newcolumntype{d}[1]{D{.}{.}{#1}}
\colorlet{tableoverheadcolor}{gray!37.5}
\colorlet{tableheadcolor}{gray!25}
\colorlet{tablerowcolor}{gray!12.5}

\def\draftdate{\relax}
\def\mda{\relax}
\def\mua{\relax}
\def\mla{\relax}
\def\draft{
\def\thtystars{******************************}
\def\sixtystars{\thtystars\thtystars}
\typeout{}
\typeout{\sixtystars**}
\typeout{* Draft mode!
         For final version remove \protect\draft\space in source file *}
\typeout{\sixtystars**}
\typeout{}
\def\draftdate{\today}
\def\mua{\marginpar[\boldmath\hfil$\uparrow$]%
                   {\boldmath$\uparrow$\hfil}\color{black}%
                    \typeout{marginpar: $\uparrow$}\ignorespaces}
\def\mda{\color{red}\marginpar[\boldmath\hfil$\downarrow$]%
                   {\boldmath$\downarrow$\hfil}%
                    \typeout{marginpar: $\downarrow$}\ignorespaces}
\def\mla{\marginpar[\boldmath\hfil$\rightarrow$]%
                   {\boldmath$\leftarrow $\hfil}%
                    \typeout{marginpar: $\leftrightarrow$}\ignorespaces}
\def\Mua{\marginpar[\boldmath\hfil$\Uparrow$]%
                   {\boldmath$\Uparrow$\hfil}\color{black}%
                    \typeout{marginpar: $\uparrow$}\ignorespaces}
\def\Mda{\color{red}\marginpar[\boldmath\hfil$\Downarrow$]%
                   {\boldmath$\Downarrow$\hfil}%
                    \typeout{marginpar: $\downarrow$}\ignorespaces}
\def\Mla{\marginpar[\boldmath\hfil\textcolor{red}{$\Rightarrow$}]%
                   {\boldmath\textcolor{red}{$\Leftarrow $}\hfil}%
                    \typeout{marginpar: $\leftrightarrow$}\ignorespaces}
\overfullrule 5pt
\oddsidemargin 15mm
\marginparwidth 29mm
}

\newcommand{\mc}{\mathcal}

\newcommand{\noun}[1]{{\scshape #1}}

\newcommand{\POWHEGBOXRES}{\noun{Powheg-Box-Res}}
\newcommand{\recolaone}{\noun{Recola}~1}
\newcommand{\recolatwo}{\noun{Recola}~2}
\newcommand{\SMEFTNLO}{\noun{Smeft@Nlo}}
\newcommand{\MadGraph}{\noun{Mg5\_aMC@Nlo}}

\DeclareSIUnit{\gev}{\giga \electronvolt}
\DeclareSIUnit{\barn}{b}

\begin{document}

\title{\hfill ~\\[-30mm]
\phantom{h}\hfill\mbox{\small COMETA-2025-27, MPP-2025-136, LAPTH-027/25}\\[1cm]
\vspace{13mm}\textbf{Polarized-boson pairs at NLO in the SMEFT}}
\date{} 
\author{
Ulrich Haisch$^{\,1\,}$\footnote{E-mail:
\texttt{haisch@mpp.mpg.de}},
Jakob Linder$^{\,1,2\,}$\footnote{E-mail:
\texttt{linder@mpp.mpg.de}}, Giovanni Pelliccioli$^{\,3\,}$\footnote{E-mail: \texttt{giovanni.pelliccioli@unimib.it}},\\
Emanuele Re$^{\,3\,}$\footnote{E-mail: \texttt{emanuele.re@mib.infn.it}}~\footnote{On leave of absence from LAPTh, CNRS - USMB, 74940 Annecy, France.}, and 
Giulia Zanderighi$^{\,1,2\,}$\footnote{E-mail: \texttt{zanderi@mpp.mpg.de}}
\\[9mm]
{\small\it $^1$ Max-Planck-Institut f\"ur Physik}\\
{\small \it Boltzmannstrasse 8, 85748 Garching, Germany}\\[3mm]
{\small\it $^2$ Technische Universit\"at M\"unchen}\\
{\small \it James-Franck-Strasse 1, 85748 Garching, Germany}\\[3mm]
{\small\it $^3$ Universit\`a degli Studi di Milano-Bicocca and INFN Sezione di Milano-Bicocca}\\
{\small \it Piazza della Scienza 3, 20126 Milano, Italy}\\[3mm]
}

\maketitle

\begin{abstract}
\noindent
We present a computation of diboson production in the $W^\pm Z$ channel at the Large Hadron Collider~(LHC), incorporating leptonic decays of the gauge bosons and considering intermediate gauge bosons with definite polarization states. The analysis includes contributions from the Standard Model effective field theory~(SMEFT) and is carried out at next-to-leading order accuracy in QCD, matched to a parton-shower simulation. Our implementation allows for the selection of specific helicity configurations, both in the Standard Model and in the presence of dimension-six operators inducing anomalous triple-gauge-boson couplings. This work provides a key ingredient for both polarization-template and quantum-tomography analyses of diboson systems at the LHC within the SMEFT framework.
\end{abstract}
\thispagestyle{empty}
\vfill
\newpage
\setcounter{page}{1}

\hrule
\tableofcontents
\hspace*{2cm}\hrule

\section{Introduction} \label{sec:intro}

A key feature of diboson production is the polarization of the resulting electroweak~(EW) gauge bosons. The $W$ and $Z$~bosons can be produced in longitudinal, right-, or left-handed states, with their relative fractions precisely predicted in the Standard Model~(SM) by its gauge structure and symmetry-breaking dynamics. The longitudinal modes, in particular, are closely tied to the Higgs sector and EW symmetry breaking~(EWSB), making them especially sensitive to new physics. Deviations in their production rates or angular distributions may signal physics beyond the~SM~(BSM). Indeed, many BSM scenarios --- such as composite Higgs models or theories with extra dimensions --- predict enhancements of the longitudinal polarization states. Within the SM effective field theory~(SMEFT)~\cite{Buchmuller:1985jz,Grzadkowski:2010es,Brivio:2017vri,Isidori:2023pyp}, such effects can be systematically analyzed via higher-dimensional operators that modify the triple- and quartic-gauge-boson couplings.

Despite the inherent challenges, polarization fractions of the EW gauge bosons have already been successfully extracted from LHC~Run~2 data~\cite{ATLAS:2019bsc,CMS:2020etf,CMS:2021icx,ATLAS:2022oge,ATLAS:2023zrv,ATLAS:2024qbd,ATLAS:2025wuw}, with improved results anticipated from LHC~Run~3 and the high-luminosity phase of the LHC~(HL-LHC). The current analysis strategies rely on template fits to data, using separate templates for each polarization component of the signal. This approach is made possible by defining polarized cross~section --- i.e., those corresponding to fixed polarizations of the intermediate gauge bosons --- which can be compared to dedicated Monte Carlo~(MC) simulations. This method enables a detailed understanding of the dynamics of diboson processes, capturing effects such as off-shell contributions and spin interference. Reaching the required level of precision demands a detailed and reliable theoretical description of both the production and decay stages of polarized gauge bosons. To address this issue, a gauge-invariant definition of polarized-boson signals has been developed~\cite{Ballestrero:2017bxn,Ballestrero:2019qoy,BuarqueFranzosi:2019boy,Ballestrero:2020qgv}. SM predictions within the pole approximation~\cite{Stuart:1991cc,Stuart:1991xk,Aeppli:1993cb,Aeppli:1993rs,Denner:2000bj,Denner:2005fg,Denner:2019vbn} and the narrow-width approximation~(NWA)~\cite{Richardson:2001df,Uhlemann:2008pm,Artoisenet:2012st} have been computed with increasing precision, reaching next-to-leading order~(NLO)~\cite{Denner:2020bcz,Denner:2020eck} and next-to-next-to-leading order~(NNLO) QCD accuracy~\cite{Poncelet:2021jmj}, including NLO~EW corrections~\cite{Denner:2021csi,Le:2022lrp,Le:2022ppa,Denner:2023ehn,Dao:2023pkl,Dao:2023kwc,Dao:2024ffg,Denner:2024tlu,Grossi:2024jae}, and NLO~QCD calculations matched to a parton shower~(PS)~\cite{Hoppe:2023uux,Pelliccioli:2023zpd}. A very recent study \cite{Carrivale:2025mjy} presents a detailed comparison of state-of-the-art predictions from various MC tools for polarized
$Z$-boson pair production, achieving for the first time a combination of NNLO QCD and NLO EW corrections. Readers interested in further details on SM calculations involving polarized diboson production at the LHC are referred to this work.

The synergy between precise measurements and high-accuracy SM predictions makes diboson processes a powerful probe of SMEFT effects. Consequently, numerous theoretical studies have explored this potential~\cite{Degrande:2012wf,Falkowski:2015jaa,Falkowski:2016cxu,Helset:2017mlf,Baglio:2017bfe,Azatov:2017kzw,Panico:2017frx,Franceschini:2017xkh,Chiesa:2018lcs,Liu:2018pkg,Grojean:2018dqj,Baglio:2018bkm,Azatov:2019xxn,Baglio:2019uty,Baglio:2020oqu,Ellis:2020unq,Degrande:2021zpv,Degrande:2023iob,Aoude:2023hxv,Degrande:2024bmd,ElFaham:2024uop,Banerjee:2024eyo,Thomas:2024dwd}, highlighting the sensitivity to both triple-gauge-boson couplings and fermion-EW-boson interactions, the latter offering complementary insight to that from EW precision measurements. One of the main challenges in this context is the suppression of the interference between the SM and certain dimension-six SMEFT operators. This is a result of helicity selection rules, which dictate that the SM and the SMEFT tree-level amplitudes for $2 \to 2$ processes are dominated by different helicity configurations~\cite{Azatov:2016sqh}. This has spurred the development of ``interference-resurrecting'' strategies, including the use of chirality-sensitive azimuthal angles of decay products~\cite{Azatov:2017kzw,Panico:2017frx,Franceschini:2017xkh,Azatov:2019xxn}, associated jet production~\cite{Azatov:2017kzw}, NLO QCD corrections~\cite{Panico:2017frx,Franceschini:2017xkh,Azatov:2019xxn}, fiducial lepton cuts~\cite{Azatov:2019xxn}, and off-shell modeling beyond the NWA~\cite{Helset:2017mlf}. Motivated by the role of chirality-sensitive observables in restoring the SM-SMEFT interference, also polarization-sensitive observables have recently been studied~\cite{ElFaham:2024uop} to enhance the sensitivity of diboson processes to SMEFT effects.

This work aims to extend and generalize the recent SMEFT analysis~\cite{ElFaham:2024uop}, which reached NLO plus PS~(NLO+PS) accuracy for two dimension-six SMEFT operators involving purely gauge-boson interactions. In this article, we examine the complete set of eight CP-even and -odd primary operators that modify both the gauge-boson self-interactions as well as the Higgs-gauge-boson couplings. The required fixed-order SMEFT amplitudes are computed using a modified version of the \recolatwo{} amplitude generator~\cite{Denner:2017vms,Denner:2017wsf}, which allows for the selection of specific helicity configurations of the intermediate gauge bosons. The PS matching is performed within the~\POWHEGBOXRES{} framework~\cite{Nason:2004rx,Frixione:2007vw,Alioli:2010xd,Jezo:2015aia}, utilizing the pole approximation. This approach enables a realistic and exclusive modeling of diboson production and their leptonic decays at NLO QCD accuracy. We~showcase the performance of our new MC code through a computation of inclusive $W^\pm Z$ production at the LHC, accounting for leptonic decays and polarized intermediate gauge bosons. Specifically, we conduct an in-depth phenomenological analysis of polarized signals, both inclusively and with realistic fiducial cuts, to quantify the magnitude of interference between the SMEFT and SM in each case. Our~study tries to identify the classes of observables in $W^\pm Z$ production most sensitive to anomalous triple-gauge-boson interactions from the considered dimension-six SMEFT operators. The obtained results offer valuable input for experimental efforts to reduce modeling uncertainties in LHC~Run~3 and HL-LHC polarization analyses of diboson processes. In particular, template-based model-independent new-physics searches stand to benefit from the new~\POWHEGBOXRES{} implementation of SMEFT~effects in diboson processes.

The structure of this article is as follows: \cref{sec:calculation} outlines the main ingredients of the SMEFT computation and their implementation within our NLO+PS event generator. In \cref{sec:anatomy}, we discuss the primary constraints on the relevant Wilson coefficients, which motivate the SMEFT benchmark scenarios considered in this work. \cref{sec:results} presents a phenomenological analysis of SMEFT corrections to $W^\pm Z$ production at the LHC, based on realistic benchmark scenarios for the Wilson coefficients. We summarize our findings and discuss future directions in~\cref{sec:conclusion}. The relevant input parameters needed to simulate polarized-boson events in the SM and in the SMEFT with the new \POWHEGBOXRES{} implementation are listed in~\cref{app:MC}.
 
\section{Details of the calculation} \label{sec:calculation}

This section begins by defining the relevant SMEFT operators considered in our work. We then review the definition of polarized signals in diboson production and decay at the LHC. Next, we detail the technical modifications made to incorporate the SMEFT operators into the \recolatwo{} amplitude generator. Finally, we outline the changes required to consistently match fixed-order~NLO~QCD predictions to a PS within the \POWHEGBOXRES{} framework, in the context of polarized diboson production with leptonic decays.

\subsection{SMEFT operators} \label{sec:framework}

To set our notation and conventions, we start by defining the SMEFT Lagrangian:
\beq \label{eq:LSMEFT}
{\cal L}_{\rm SMEFT}= \sum_i \frac{C_i (\mu)}{\Lambda^2}\hspace{0.5mm}Q_i \,.
\eeq
Here, $C_i (\mu)$ represents dimensionless Wilson coefficients evaluated at the renormalization scale $\mu$, which multiply the corresponding effective operators $Q_i$. Throughout this article, we assume that all Wilson coefficients are real. The symbol $\Lambda$ denotes the common new-physics scale suppressing the operators. 

The dimension-six operators we consider in this article are expressed in the Warsaw operator basis~\cite{Grzadkowski:2010es} as follows:
\beq \label{eq:operators} 
\begin{gathered} 
Q_{HB} = H^{\dagger} H \hspace{0.25mm} B_{\mu \nu} B^{\mu \nu}\,, \qquad
Q_{H\widetilde{B}} = H^{\dagger} H \hspace{0.25mm} B_{\mu \nu} \widetilde{B}^{\mu \nu}\,, \\[2mm]
Q_{HW} = H^{\dagger} H \hspace{0.25mm} W_{\mu \nu}^i W^{i, \mu \nu}\,,
\qquad
Q_{H\widetilde{W}} = H^{\dagger} H \hspace{0.25mm} W_{\mu \nu}^{i} \widetilde{W}^{i, \mu \nu}\,, \\[2mm]
Q_{HW\!B} = H^{\dagger} \sigma^{i} H \hspace{0.25mm} W^{i}_{\mu\nu} B^{\mu\nu}\,,
\qquad
Q_{H\widetilde{W}\!B} = H^{\dagger} \sigma^{i} H \hspace{0.25mm} \widetilde{W}^{i}_{\mu\nu} B^{\mu\nu}\,, \\[2mm]
Q_W = \epsilon_{ijk} \hspace{0.25mm} W^{i, \nu}_{\mu} \hspace{0.25mm} W^{j, \lambda}_{\nu} \hspace{0.25mm}W^{k, \mu}_{\lambda}\,,
\qquad
Q_{\widetilde{W}} = \epsilon_{ijk} \hspace{0.25mm} W^{i, \nu}_{\mu} \hspace{0.25mm} W^{j, \lambda}_{\nu} \hspace{0.25mm} \widetilde{W}^{k, \mu}_{\lambda}\,. 
\end{gathered}
\eeq
Here, $H$ represents the SM Higgs doublet, while the field-strength tensors associated with the $U(1)_Y$ and $SU(2)_L$ gauge fields are $B_{\mu \nu}= \partial_\mu B_\nu - \partial_\nu B_\mu$ and $W_{\mu \nu}^i = \partial_\mu W_\nu^i - \partial_\nu W_\mu^i - g_2 \hspace{0.125mm}\epsilon_{ijk}\hspace{0.125mm}W_\mu^j \hspace{0.125mm}W_\nu^k$, respectively. The~corresponding gauge couplings are called $g_1$ and $g_2$. The symbol $\sigma^i$ denotes the standard Pauli matrices, and $\epsilon_{ijk}$ represents the totally antisymmetric Levi-Civita symbol, defined such that $\epsilon_{123}= +1$. The dual field-strength tensors are defined as $\widetilde{B}_{\mu\nu}= \epsilon_{\mu \nu \lambda \rho}\hspace{0.25mm}B^{\lambda \rho}/2$ and $\widetilde{W}_{\mu\nu}^i = \epsilon_{\mu \nu \lambda \rho}\hspace{0.25mm}W^{i, \lambda \rho}/2$, where $\epsilon_{\mu \nu \lambda \rho}$ is the fully antisymmetric four-dimensional Levi-Civita tensor with $\epsilon_{0123}= +1$. Note that the operators in~\cref{eq:operators} without a tilde are CP-even, while those with a tilde are CP-odd. Since the gauge-boson interactions in the SM conserve the CP symmetry, only the CP-even operators can interfere with the SM contribution to diboson production, whereas the CP-odd operators do~not. We note that the recent work~\cite{ElFaham:2024uop} has provided NLO-accurate results for fully leptonic $W^\pm Z$ and $W^+ W^-$ production at the LHC, focusing on the two triple-gauge-boson operators~$Q_W$ and~$Q_{\widetilde{W}}$.

\subsection{Definition of polarized signals} \label{sec:polsig}

A gauge-invariant definition of polarized signals requires that intermediate gauge bosons be constrained to their mass shell. Consequently, the simulation of polarized-boson production and decay involves additional subtleties compared to a fully off-shell approach. In practical terms, enabling a MC generator to simulate intermediate gauge bosons with fixed polarization states demands two key elements: $(i)$ the implementation of an on-shell projection to isolate the gauge-invariant resonant contribution to the amplitude, and $(ii)$ the extraction of individual polarization states from the resonant amplitude.

\begin{figure}[!t]
\centering
\includegraphics[width=.65\textwidth]{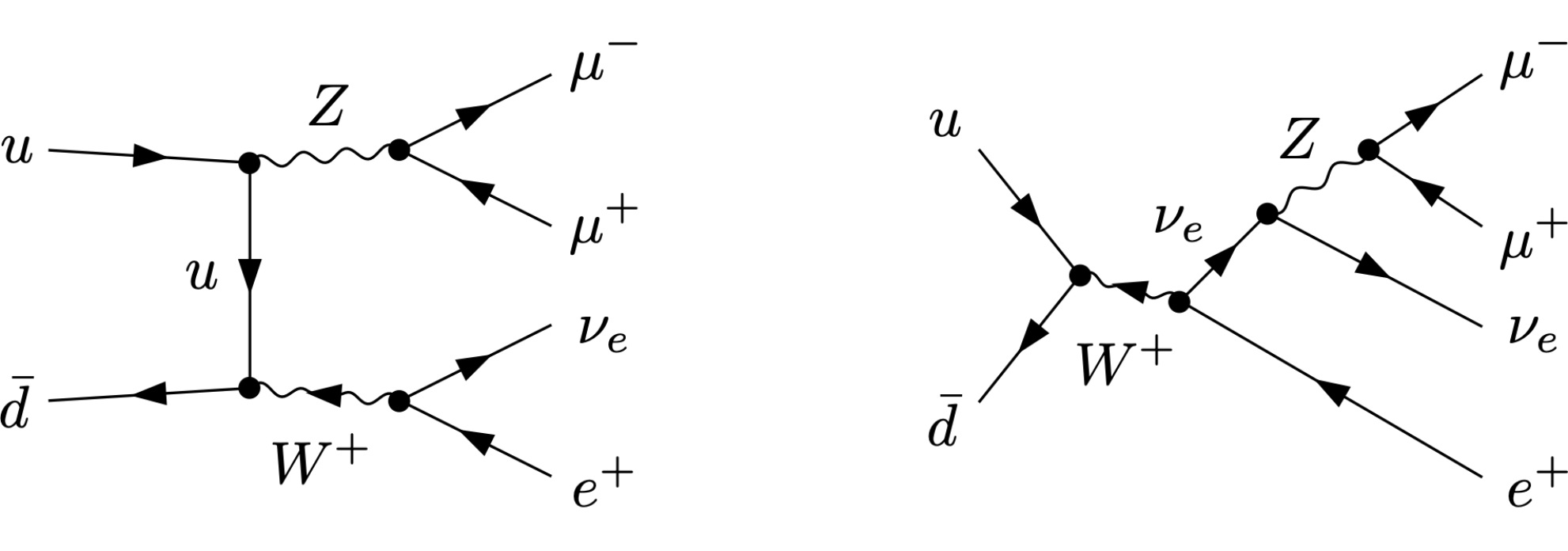}
\vspace{0mm}
\caption{Example tree-level resonant (left) and non-resonant (right) diagrams contributing to $W^+ Z$ production and decay in proton-proton ($pp$) collisions. Further explanations can be found in the main text. \label{fig:WZLO}}
\end{figure}

We illustrate the approach using the tree-level contribution to $W^+ Z$ production as a representative example. Sample SM diagrams are shown in~\cref{fig:WZLO}. At a given perturbative order, the full gauge-invariant set of diagrams includes both resonant and non-resonant contributions. In this context, we retain only the resonant diagrams, discarding the non-resonant ones. For diboson production, this corresponds to selecting diagrams that factorize into two distinct production vertices, two $s$-channel gauge-boson propagators, and separate decay vertices for each boson. Accordingly, the term “resonant” is understood to mean “doubly-resonant”. This can be expressed schematically~as:
\beq \label{eq:ampred}
\begin{split}
{\cal A}_{\rm full}(x_1, x_2; k_1, \ldots, k_4) & = {\cal A}_{\text{non-res}}(x_1, x_2; k_1, \ldots, k_4) + {\cal A}_{\rm res}(x_1, x_2; k_1, \ldots, k_4)\\[2mm]
& \longrightarrow \, {\cal A}_{\rm res}(x_1, x_2; k_1, \ldots, k_4) \,. 
\end{split}
\eeq
Here, $x_1$ and $x_2$ denote the momentum fractions of the incoming partons, and $k_1, \ldots, k_4$ correspond to the final-state four-momenta. In the ’t~Hooft-Feynman gauge, the resonant contribution to the amplitude is given by
\beq \label{eq:Ares}
\begin{split}
{\cal A}_{\rm res}(x_1, x_2; k_1, \ldots, k_4) & = {\cal P}_{\mu \nu}(x_1, x_2; k_{12}, k_{34}) \, \frac{-i \hspace{0.125mm}g^{\mu \alpha}}{k_{12}^2 - \mu_W^2}\, \frac{-i \hspace{0.125mm}g^{\nu \beta}}{k_{34}^2 - \mu_Z^2}\, {\cal D}^W_\alpha (k_1, k_2) \,{\cal D}^Z_\beta (k_3, k_4) \,,
\end{split}
\eeq
where $k_{i \ldots n}= k_i + \ldots + k_n$, and for $V = W,Z$, we defined
\beq \label{eq:muV2}
\mu_V^2 = m_V^2 - i \hspace{0.25mm}\Gamma_V \hspace{0.25mm}m_V \,, 
\eeq 
with $m_W$, $m_Z$ and $\Gamma_W$, $\Gamma_Z$ denoting the pole masses and total decay widths of the $W$ and $Z$~bosons, respectively. The tensor ${\cal P}_{\mu\nu}(x_1, x_2; k_{12}, k_{34})$ encodes the sub-amplitude for the production of the two gauge bosons, while ${\cal D}_\alpha^W (k_1, k_2)$ and ${\cal D}_\beta^Z (k_3, k_4) $ represent the decay sub-amplitudes of the relevant gauge boson.

Since the subset of resonant diagrams is not gauge invariant by itself, we ensure gauge invariance of the calculation by employing the double-pole approximation~(DPA), as also adopted for instance in~Refs.~\cite{Denner:2020bcz,Denner:2020eck,Denner:2021csi,Denner:2023ehn,Denner:2024tlu,Pelliccioli:2023zpd,Denner:2022riz}. The first step in the DPA procedure involves the following phase-space mapping
\beq \label{eq:map}
\Phi_4 = \{x_1, x_2; k_1, \ldots, k_4 \}\, \stackrel{\rm DPA}{\longrightarrow}\, \tilde{\Phi}_4 = \{x_1, x_2; \tilde{k}_1, \ldots, \tilde{k}_4 \}\,,
\eeq
with $\tilde k_{12}= m_W^2$ and $\tilde k_{34}= m_Z^2$. The mapping in~\cref{eq:map} is now applied to the resonant part of the amplitude in~\cref{eq:ampred}, such that the numerator is projected on shell, while the propagator denominators retain the original off-shell kinematics. In~schematic form, this corresponds to the following replacement: 
\beq \label{eq:Aresmapped}
\begin{split}
{\cal A}_{\rm res}(x_1, x_2; k_1, \ldots, k_4) \, \longrightarrow \, {\cal A}_{\rm res}(x_1, x_2; \tilde{k}_1, \ldots, \tilde{k}_4) & = {\cal P}_{\mu \nu}(x_1, x_2; \tilde{k}_{12}, \tilde{k}_{34}) \\[2mm] 
& \hspace{-2cm}\phantom{xx}\times \frac{-i \hspace{0.125mm}g^{\mu \alpha}}{k_{12}^2 - \mu_W^2}\, \frac{-i \hspace{0.125mm}g^{\nu \beta}}{k_{34}^2 - \mu_Z^2}\, {\cal D}^W_\alpha (\tilde{k}_1, \tilde{k}_2) \,{\cal D}^Z_\beta (\tilde{k}_3, \tilde{k}_4) \,.
\end{split}
\eeq
In the production ($\mc P$) and decay ($\mc D$) terms present on the right-hand side of~\cref{eq:Aresmapped} the widths $\Gamma_W, \Gamma_Z$ of the weak bosons are set to zero in order to ensure EW gauge invariance. Note that the mappings in~\cref{eq:map} enforce a kinematic threshold, $m_{4 \ell}^2 = k_{1234}^2 > (m_W + m_Z)^2$, on the invariant mass of the four-lepton final state to permit the production of on-shell $W$ and $Z$~bosons. However, the mapping itself is not uniquely defined. For our purposes, we adopt the on-shell mapping procedure from~\citere{Denner:2021csi}, which preserves the following quantities: $(i)$~the total four-momentum of the diboson system, $k_{1234}$; $(ii)$~the spatial direction of $k_{12}$~(and $k_{34}$) in the diboson center-of-mass~(CM) frame; $(iii)$~the spatial direction of $k_1$~(and $k_2$) in the rest frame of $k_{12}$; $(iv)$ the spatial direction of~$k_3$~(and~$k_4$) in the rest frame of $k_{34}$; $(v)$~the initial-state parton momentum fractions, $x_1$ and $x_2$; and $(vi)$~the four-momentum of any additional jet radiation present at NLO in QCD. The choice in~\citere{Denner:2021csi} simplifies the DPA implementation, particularly for subtracted-real corrections~\cite{Denner:2020bcz}. Notably, the on-shell projection preserves initial-state energy fractions and decay angles, making it well-suited for polarization studies. We note that alternative mappings have been explored in the literature, but variations in their implementation have only a minor impact on numerical results in practice~\cite{Denner:2021csi,Carrivale:2025mjy}.

Squaring the DPA-evaluated resonant contributions from~\cref{eq:Aresmapped}, and multiplying by the parton luminosities and phase-space weights, yields the unpolarized cross section (differential in any observable). To isolate a specific boson polarization mode, the Lorentz-invariant tensor structure in the propagator numerators is replaced with the corresponding polarization-vector contributions. In~the ’t~Hooft-Feynman gauge, the replacement for a single gauge boson with four-momentum $k$ takes the form
\beq \label{eq:gmurep}
-g_{\mu \nu}= \sum_{\kappa}\epsilon_\mu^{(\kappa)}(k) \, \epsilon_\nu^{(\kappa)}(k) \, \longrightarrow \, \epsilon_\mu^{(\lambda)}(k) \, \epsilon_\nu^{(\lambda)}(k) \,, \qquad \lambda = L, +, - \,,
\eeq
where the labels $L$, $+$, and $-$ denote the longitudinal, right-, and left-handed physical polarization states, respectively. Note that on the left-hand side of~\cref{eq:gmurep}, the sum over $\kappa$ runs over four polarization states: the three physical ones and an additional unphysical state. This fourth polarization term is always cancelled by the associated would-be Goldstone-boson contribution, regardless of the perturbative order. We also emphasize that the polarization vectors in~\cref{eq:gmurep} are frame-dependent, as they rely on the Lorentz frame in which the helicity states of the gauge bosons are defined. Consequently, polarized signals are calculated in a specific Lorentz frame. The most natural choice~\cite{Denner:2020eck,Denner:2021csi,Le:2022lrp,Le:2022ppa} for inclusive diboson production is the boson-pair CM frame, which is also the frame used in this work.

Before proceeding, we emphasize that to define a polarized signal beyond the LO, the DPA and polarization selection must be applied to all cross-section contributions at NLO in QCD, including the Born, the virtual, the real, and the subtraction-counterterm corrections. Since polarization vectors are defined in the boson-pair CM frame, extra care is needed in the case of the real contributions. In fact, in Born-like contributions, the diboson and partonic CM frames align, but for real contributions, they differ by a boost. Therefore, real matrix elements must be evaluated with their four-momenta boosted from the partonic CM frame (the standard frame in the \POWHEGBOXRES{} framework) to the diboson CM frame.

\subsection{\recolatwo{} interface} \label{sec:recola}

For the purposes of this study, we adapted the \recolatwo{} amplitude generator to allow for the selection of specific helicity states of the intermediate gauge bosons via the replacement defined in~\cref{eq:gmurep}. Compared to~Refs.~\cite{Denner:2020bcz,Denner:2020eck,Denner:2021csi,Denner:2023ehn,Denner:2024tlu,Pelliccioli:2023zpd,Denner:2022riz} which all rely on an interface to the \recolaone{} SM amplitudes~\cite{Actis:2016mpe}, our approach is more general, as it allows for the computation of polarized amplitudes in any model supported by \recolatwo{}. Although fermions are treated as unpolarized in the current implementation, extending the code to include polarized fermions would be straightforward. To incorporate the SMEFT operators of~\cref{eq:operators} in \recolatwo{}, we used the \noun{Feynrules} package~\cite{Alloul:2013bka} to implement the model and produced the corresponding \noun{Ufo}~output~\cite{Degrande:2011ua}. The modified \recolatwo{} amplitude generator can calculate both tree-level and one-loop SM and SMEFT amplitudes for diboson production with leptonic decays, using the \noun{Collier} library~\cite{Denner:2016kdg} for tensor-integral reduction and evaluation. The~resulting amplitudes are then passed to the \POWHEGBOXRES{} package, which functions also as the MC integrator, following the procedure described below in~Section~\ref{sec:POWHEGBOXRES}.

The new \recolatwo{} interface underwent multiple validation steps. In particular, the implementation of the operators $Q_{W}$ and $Q_{\widetilde{W}}$ in~\cref{eq:operators} within \recolatwo{} was validated by comparing it to the results of~\citere{Chiesa:2018lcs}, which are based on the \noun{Sm+Atgc} model~\cite{SMATGC}, and found to be in complete agreement. We further validated our BSM model implementation of these operators by successfully comparing the cross-section predictions for $W^+ Z$ production with those obtained using the updated version of \SMEFTNLO{} \cite{Degrande:2020evl}, which relies on \MadGraph{}~\cite{Alwall:2014hca} for MC generation. This combination of codes was also employed in the recent article~\cite{ElFaham:2024uop}. The differential and integrated results for this validation step are shown later in~\cref{sec:validation}. The correctness of the polarization selection was confirmed through two methods. First, by comparing the complex amplitudes involving polarized gauge bosons with factorized expressions --- specifically, by generating a pair of polarized gauge bosons and multiplying their production amplitudes with the corresponding decay amplitudes. Second, by applying a Legendre-projection procedure (see, for example,~\citere{Ballestrero:2017bxn} for details), which allows for the extraction of the polarization fractions of the intermediate gauge bosons. Both tests of the polarization selection in the new \recolatwo{} interface were successful. The differential results obtained from the Legendre-projection procedure are presented below in~\cref{sec:wpol}.

\subsection{\POWHEGBOXRES{} implementation} \label{sec:POWHEGBOXRES}

In the \POWHEGBOXRES{} framework, the QCD soft and collinear singularities are subtracted using the Frixione-Kunszt-Signer~(FKS) scheme~\cite{Frixione:2007vw,Jezo:2015aia,Frixione:1995ms}. It is therefore essential that the DPA procedure maintains the local cancellation between real contributions and subtraction counterterms in the radiative phase space $\Phi_{\rm rad}$. This is accomplished by first applying the FKS-subtraction mapping, followed by the DPA on-shell mapping, to evaluate the subtracted-real kinematics. Schematically, one has for the final-state four-lepton phase space
\beq \label{eq:FKSDPA}
\begin{split}
\Phi_4 = \{x_1, x_2; k_1, \ldots, k_4 \}& \, \stackrel{\rm FKS}{\longrightarrow}\, \{\bar{\Phi}_4, \Phi_{\rm rad}\}= \{\bar{x}_1, \bar{x}_2; \bar{k}_1, \ldots, \bar{k}_4, k_{\rm rad}\}\\[2mm] 
& \, \stackrel{\rm DPA}{\longrightarrow}\, \{\tilde{\bar{\Phi}}_4, \Phi_{\rm rad}\}= \{\bar{x}_1, \bar{x}_2; \tilde{\bar{k}}_1, \ldots, \tilde{\bar{k}}_4, k_{\rm rad}\}\,,
\end{split}
\eeq
where $k_{\rm rad}$ denotes the four-momentum of the radiation. Note that in the case of QCD initial-state radiation --- the only type of QCD radiation present in the processes considered --- the cancellation of infrared (IR) singularities in the real phase space is ensured by our specific choice of DPA mapping $\big($see the discussion following~\cref{eq:Aresmapped}$\big)$. This mapping preserves the total momentum of the color-singlet system and thus commutes with the FKS mapping, which applies a boost uniformly to all four final-state leptons. As a result, the subtraction of IR singularities proceeds analogously to the full off-shell calculation. 

The preceding discussion shows that, modulo minor technical modifications, the \POWHEGBOXRES{} matching procedure can be implemented analogously to the off-shell computation of diboson production with leptonic decays. In particular, the standard matching formula, adapted to such processes within the DPA framework, takes the form:
\beq \label{eq:matching}
\langle {\cal O}\rangle = \int \! d\Phi_4 \, \bar{B}(\tilde{\Phi}_4) \left [{\cal O}(\tilde{\Phi}_4) \, \Delta (t_0) + \int_{t>t_0}\! d \Phi_{\rm rad}\, {\cal O}(\tilde{\bar{\Phi}}_4, \Phi_{\rm rad}) \, \frac{R (\tilde{\bar{\Phi}}_4, \Phi_{\rm rad})}{B (\tilde{\Phi}_4)}\, \Delta (t) \right ] \,.
\eeq
Here, the Sudakov form factor $\Delta (t)$ is given by
\beq \label{eq:sudakov}
\Delta (t) = \exp \left [ - \int_{t^\prime > t}\! d\Phi_{\rm rad}^\prime \, \frac{R (\tilde{\bar{\Phi}}_4, \Phi_{\rm rad}^\prime)}{B (\tilde{\Phi}_4)}\right ] \,,
\eeq
and the ordering variable $t$ is the transverse momentum of the radiation. The NLO-weighted Born term $\bar{B}(\tilde{\Phi}_4)$ takes the form:
\beq \label{eq:Bbar}
\bar{B}(\tilde{\Phi}_4) = B (\tilde{\Phi}_4) + V_{\rm reg}(\tilde{\Phi}_4) + \int \! d\Phi_{\rm rad}\, \left [ R (\tilde{\bar{\Phi}}_4, \Phi_{\rm rad}) - C (\tilde{\bar{\Phi}}_4, \Phi_{\rm rad}) \right ] \,.
\eeq
Here, $B (\tilde{\Phi}_4)$ denotes the squared Born-level matrix element, $V_{\rm reg}(\tilde{\Phi}_4)$ represents the finite virtual corrections together with the integrated FKS subtraction counterterms (including collinear remnants), $R (\tilde{\bar{\Phi}}_4, \Phi_{\rm rad})$ corresponds to the real-emission matrix element squared, and $C (\tilde{\bar{\Phi}}_4, \Phi_{\rm rad})$ is the associated local subtraction counterterm. In~\cref{eq:matching}, it is understood that the DPA-mapped kinematics --- $\tilde{\bar{\Phi}}_4$ for Born-like terms and $\{\tilde{\bar{\Phi}}_4, \Phi_{\rm rad}\}$ for real-emission contributions --- are employed in the matrix-element numerators. In contrast, the original off-shell kinematics --- $\Phi_4$ and $\{\bar{\Phi}_4, \Phi_{\rm rad}\}$, respectively --- are retained in the resonant matrix-element denominators and phase-space weights. Likewise, any selection cuts are consistently applied to the off-shell kinematics.

Compared to the off-shell computation, the main subtlety in the computation of the resonant contributions lies in the kinematic constraint imposed by the DPA, which for $W^+ Z$ production with leptonic decays requires the invariant mass of the final-state four-lepton system to satisfy $m_{4 \ell}^2 = k_{1234}^2 > (m_W + m_Z)^2$. In the \POWHEGBOXRES{} framework, where real-emission kinematics is constructed from Born-like configurations, phase-space points that violate this condition are simply rejected by assigning a zero MC Jacobian to them. This guarantees numerical stability in computing the ratio $R (\tilde{\bar{\Phi}}_4, \Phi_{\rm rad})/B (\tilde{\Phi}_4)$ in the matching formula~\cref{eq:matching}.

\section{Anatomy of SMEFT effects} \label{sec:anatomy}

In this section, we motivate a set of simple SMEFT benchmark scenarios by examining the leading experimental constraints on the Wilson coefficients associated with the dimension-six operators defined in~\cref{eq:operators}. The analyses presented in Sections~\ref{sec:wbosonmass} to~\ref{sec:edms} are performed using the LEP~scheme, in which the weak mixing angle $\theta_w$, the gauge couplings $g_1$ and $g_2$, and the Higgs vacuum expectation value $v$ are expressed in terms of the EW input parameters $\alpha$, $G_F$, and $m_Z$. Here, $\alpha$ is the electromagnetic coupling constant, $G_F$ is the Fermi constant extracted from muon decay, and $m_Z$ is the mass of the $Z$ boson. Beginning in Section~\ref{sec:TGCs}, we also introduce the $G_F$~scheme, which is the convention adopted for the new \POWHEGBOXRES{} implementation. For additional technical details on EW input schemes within the SMEFT framework, we refer the interested reader to Refs.~\cite{Brivio:2017vri,Brivio:2020onw,Biekotter:2023xle,Gauld:2023gtb}.

\subsection[$W$-boson mass]{$\boldsymbol{W}$-boson mass} \label{sec:wbosonmass}

In the LEP scheme, the mass of the $W$~boson is a predicted quantity. At tree level within the~SM, the relation $m_W = c_w \hspace{0.25mm}m_Z$ holds, where $c_w \simeq 0.88$ is the cosine of the weak mixing angle, and $m_Z \simeq 91.2 \, {\rm GeV}$. The~relative modification of $m_W$ arising from the SMEFT operators specified in~\cref{eq:operators} is expressed~as
\beq \label{eq:dMWoMWtheo}
\begin{split}
\frac{\delta m_W}{m_W}& \simeq -\frac{c_w \hspace{0.25mm}s_w}{c_w^2 - s_w^2}\frac{v^2}{\Lambda^2}\left \{C_{HW\!B}- \frac{\alpha}{4 \pi}\frac{1}{c_w \hspace{0.25mm}s_w}\left [ C_{HB}+ C_{HW}+ \frac{3 e}{2 s_w}\hspace{0.5mm}C_W \right ] \ln \left ( \frac{\Lambda^2}{m_W^2}\right ) \right \}\,,
\end{split}
\eeq
where $s_w \simeq 0.48$ is the sine of the weak mixing angle, $v \simeq 246 \, {\rm GeV}$, $\alpha \simeq 1/128$, and $e =\sqrt{4 \pi \alpha}$. The~first term in~\cref{eq:dMWoMWtheo} reflects the tree-level contribution, while the subsequent terms account for the logarithmically enhanced corrections generated by one-loop renormalization group~(RG) evolution in the SMEFT from the scale $\Lambda$ down to $m_W$. Accordingly, the Wilson coefficients in~\cref{eq:dMWoMWtheo} are to be understood as evaluated at the scale~$\Lambda$. It is important to note that the logarithmically enhanced corrections can be obtained from the results in~\citere{Alonso:2013hga}, whereas computing the finite contributions associated with $C_{HB}$, $C_{HW}$, and $C_W$ necessitates a complete one-loop calculation~\cite{Dawson:2019clf,Biekotter:2025nln}. As we aim to illustrate the structure of the SMEFT corrections to the $W$-boson mass resulting from~\cref{eq:operators}, it is sufficient to focus solely on the leading logarithmic~(LL) one-loop effects in~\cref{eq:dMWoMWtheo} as far as quantum corrections are concerned. Similarly, the LL contribution proportional to $C_{HW\!B}$ is omitted from~\cref{eq:dMWoMWtheo}, since this Wilson coefficient already contributes at tree level.

Based on the latest world average of $m_W$ from the Particle~Data~Group~(PDG)~\cite{LHC-TeVMWWorkingGroup:2023zkn,ParticleDataGroup:2024cfk}, excluding the CDF~II measurement~\cite{CDF:2022hxs}, and the most advanced SM prediction~\cite{Chen:2020xot}, we obtain the following $95\%$ confidence level~(CL) limit
\beq \label{eq:dMWoMWexp}
\frac{\delta m_W}{m_W}\in [-5.1, 1.8] \cdot 10^{-4}\,,
\eeq
on the allowed relative shift of $m_W$. This result imposes relevant constraints on the Wilson coefficients appearing in \cref{eq:dMWoMWtheo}. As an example, for the operator $Q_{HW\!B}$, we obtain the following bound
\beq \label{eq:CHWBbound}
\frac{C_{HW\!B}}{\Lambda^2}\in [-0.4, 1.1] \cdot 10^{-2}\, {\rm TeV}^{-2}\,,
\eeq
if all other Wilson coefficients in \cref{eq:dMWoMWtheo} are set to zero. The remaining Wilson coefficients are subject to notably weaker limits owing to their loop-level suppression. For example, we find
\beq \label{eq:CHBbound}
\frac{C_{HB}}{\Lambda^2}\in [-1.1, 0.4] \, {\rm TeV}^{-2}\,,
\eeq
with the same (a slightly weaker) one-parameter bound applying to $C_{HW}$ ($C_W$). This constraint is derived by setting $\Lambda = 2 \, {\rm TeV}$ in~\cref{eq:dMWoMWtheo}.

\subsection{Higgs-boson observables} \label{sec:Higgsphysics}

The Wilson coefficients of the operators introduced in~\cref{eq:operators} affect the Higgs signal strengths in channels with two gauge bosons in the final state. The strongest constraints comes from the $h \to \gamma \gamma$ decay mode. The signal strength modification relative to the SM in the $h \to \gamma \gamma$ decay takes the~form:
\beq \label{eq:digammasignalstrength}
\begin{split}
\delta \mu_{\gamma \gamma}& \simeq \frac{2}{g_{h\gamma\gamma}}\frac{v^2}{\Lambda^2}\, \left [ c_w^2 \hspace{0.25mm}C_ {HB}+ s_w^2 \hspace{0.25mm}C_{HW}- c_w \hspace{0.25mm}s_w \hspace{0.25mm}C_{HW\!B}+ \frac{\alpha}{4 \pi}\hspace{0.25mm}\frac{9 e}{s_w}\hspace{0.5mm}C_W \ln \left ( \frac{\Lambda^2}{m_W^2}\right )\right ]\\[2mm]
& \phantom{xx}+ \frac{1}{g_{h\gamma\gamma}^2}\frac{v^4}{\Lambda^4}\, \left [ c_w^2 \hspace{0.25mm}C_ {H\widetilde{B}}+ s_w^2 \hspace{0.25mm}C_{H\widetilde{W}}- c_w \hspace{0.25mm}s_w \hspace{0.25mm}C_{H\widetilde{W}\!B}+ \frac{\alpha}{4 \pi}\hspace{0.25mm}\frac{9 e}{s_w}\hspace{0.5mm}C_{\widetilde{W}}\ln \left ( \frac{\Lambda^2}{m_W^2}\right )\right ]^2 \,. 
\end{split}
\eeq
Here, $g_{h\gamma\gamma}\simeq -1.88 \cdot 10^{-3}$ characterizes the loop-induced $h\gamma\gamma$ coupling in the SM. Explicit analytical expressions for $g_{h\gamma\gamma}$ can be found, for example, in~\citere{Brivio:2020onw}. As in the case of~\cref{eq:dMWoMWtheo}, the expression in~\cref{eq:digammasignalstrength} includes the tree-level contributions from the Higgs-gauge-boson operators, as well as the LL one-loop effects associated to the triple-gauge-boson operators. The logarithmically enhanced corrections can once again be computed using the one-loop anomalous dimensions provided in~\citere{Alonso:2013hga}, while the full one-loop corrections proportional to $C_W$ have been calculated in~\citere{Dedes:2018seb}. LL~effects proportional to Wilson coefficients other than $C_W$ and $C_{\widetilde{W}}$ are omitted from~\cref{eq:digammasignalstrength}, since they already contribute at tree level. Finally, note that the CP-even operators interfere with the SM, yielding contributions of ${\cal O}(v^2/\Lambda^2)$, whereas the CP-odd operators do not interfere and thus contribute first at ${\cal O}(v^4/\Lambda^4)$.

At the $95\%$~CL, the relative shift in the corresponding Higgs signal strength is
\beq \label{eq:haameasured}
\delta \mu_{\gamma \gamma}\in [-0.13, 0.23] \,,
\eeq
which represents an unofficial weighted average of the ATLAS~\cite{ATLAS:2020qdt} and CMS~\cite{CMS:2020gsy} measurements. The strongest single-parameter bounds that derive from~\cref{eq:haameasured} are 
\beq \label{eq:CHBbounds}
\frac{C_{HB}}{\Lambda^2}\in [-4.6, 2.6] \cdot 10^{-3}\, {\rm TeV}^{-2}\,,
\qquad \frac{C_{H\widetilde{B}}}{\Lambda^2}\in [-1.9, 1.9] \cdot 10^{-2}\, {\rm TeV}^{-2}\,,
\eeq
while the weakest limits read 
\beq \label{eq:CWbounds}
\frac{C_{W}}{\Lambda^2}\in [-1.5, 0.8] \cdot 10^{-1}\, {\rm TeV}^{-2}\,, \qquad \frac{C_{\widetilde{W}}}{\Lambda^2}\in [-6.3, 6.3] \cdot 10^{-1}\, {\rm TeV}^{-2}\,,
\eeq
owing to the loop suppression of the $Q_W$ and $Q_{\widetilde{W}}$ contributions. The loop-level bounds once again use $\Lambda = 2 \, {\rm TeV}$ as the new-physics scale in the logarithmic terms. Slightly weaker bounds than those presented in~\cref{eq:CHBbounds} apply to $C_{HW}$ and $C_{HW\!B}$, and their CP-odd counterparts. Also note that the bounds on the CP-conserving operators are generally stronger than those on the CP-violating operators, since the latter do not interfere with the SM $h \to \gamma \gamma$ amplitude. 

\subsection{Electric dipole moments} \label{sec:edms}

The CP-violating operators in~\cref{eq:operators} induce electric dipole moments~(EDMs) for the SM fermions at the loop level. The logarithmically enhanced corrections can again be determined from~Refs.~\cite{Alonso:2013hga,Jenkins:2013wua} by analyzing the RG flow of the relevant operators within the SMEFT framework. In the case of the electron EDM, we obtain (see also~\citere{Panico:2018hal}) 
\beq \label{eq:deovere}
\begin{split}
\frac{d_e}{e}& \simeq \frac{1}{16 \hspace{0.125mm}\pi^2}\frac{y_e v}{\sqrt{2}\Lambda^2}\left ( 3 \hspace{0.25mm}C_{H\widetilde{B}}+ C_{H\widetilde{W}}- \frac{3}{2 \hspace{0.125mm}c_w \hspace{0.25mm}s_w}\hspace{0.5mm}C_{H\widetilde{W}\!B}\right ) \ln \left ( \frac{\Lambda^2}{m_W^2}\right ) \\[2mm]
& \phantom{xx}+ \frac{1}{256 \hspace{0.125mm}\pi^4}\frac{y_e v}{\sqrt{2}\Lambda^2}\frac{3 e^3 \left ( 13 \hspace{0.125mm}c_w^2 + 3 \hspace{0.125mm}s_w^2 \right )}{8 \hspace{0.125mm}c_w^2 \hspace{0.25mm}s_w^3}\hspace{0.5mm}C_{\widetilde{W}}\hspace{0.25mm}\ln^2 \left ( \frac{\Lambda^2}{m_W^2}\right ) \,,
\end{split}
\eeq
where $y_e = \sqrt{2}m_e/v$ represents the Yukawa coupling of the electron, with $m_e \simeq 511 \, {\rm keV}$ being the electron mass. Note that the Wilson coefficients $C_{H\widetilde{B}}$, $C_{H\widetilde{W}}$, and $C_{H\widetilde{W}\!B}$ appear~in~\cref{eq:deovere} at the one-loop level, while $C_{\widetilde{W}}$ contributes only at two loops. The expression shown includes only the LL corrections, which, for instance, can be straightforwardly extracted from the formulas given in~\citere{Haisch:2025lvd}. The single-logarithmic contributions to $d_e/e$ arise from the one-loop mixing of $Q_{H\widetilde{B}}$, $Q_{H\widetilde{W}}$, and~$Q_{H\widetilde{W}\!B}$ into the EW dipole operators 
\beq \label{eq:EWdipole}
Q_{eB}= \left ( L \hspace{0.25mm}\sigma^{\mu \nu}e \right ) H \hspace{0.25mm}B_{\mu \nu}\,, \qquad Q_{eW}= \left ( L \hspace{0.25mm}\sigma^i \sigma^{\mu \nu}e \right ) H \hspace{0.25mm}W_{\mu \nu}^i \,, 
\eeq
involving the left-handed first-generation lepton field $L$ and the right-handed electron field~$e$. Here, $\sigma_{\mu \nu} = i \hspace{0.5mm} [\gamma_\mu, \gamma_\nu]/2$ with $\gamma_\mu$ the usual Dirac matrices and the square brackets denoting the commutator. In contrast, the double-logarithmic terms originate from the one-loop mixing of $Q_{\widetilde{W}}$ into these intermediate operators, which then mix into the dipole operators~\cref{eq:EWdipole}. Two-loop effects associated with the Wilson coefficients $C_{H \widetilde{B}}$, $C_{H \widetilde{W}}$, and $C_{H \widetilde{W}\!B}$ are omitted from~\cref{eq:deovere}.

The most stringent upper bound on the electron EDM has been established in~\citere{Roussy:2022cmp}. At the $90\%$~CL, one has:
\beq \label{eq:expeEDM}
\left | \frac{d_e}{e}\right | < 4.1 \cdot 10^{-30}\, {\rm cm}\,.
\eeq
The strongest constraint on the CP-odd Wilson coefficients contributing to~\cref{eq:deovere} at the one-loop level is given by
\beq \label{eq:CHWtildeBbound}
\frac{\left | C_{H\widetilde{W}\!B}\right |}{\Lambda^2}< 2.8 \cdot 10^{-6}\, {\rm TeV}^{-2}\,. 
\eeq
Slightly weaker bounds apply to the other CP-odd Higgs-gauge-boson operators. The $90\%$~CL bound on the Wilson coefficient of the operator $Q_{\widetilde{W}}$, which contributes at the two-loop level, instead reads: 
\beq \label{eq:CHWtildebound}
\frac{\left | C_{\widetilde{W}}\right |}{\Lambda^2}< 5.1 \cdot 10^{-4}\, {\rm TeV}^{-2}\,.
\eeq
The stated bounds assume a scale of $\Lambda = 2 \, {\rm TeV}$ for the logarithmic terms in the expression for~$d_e/e$. Observe that the constraints in~\cref{eq:CHWtildeBbound,eq:CHWtildebound} are
considerably stronger than those on the CP-violating operators given in~\cref{eq:CHBbounds,eq:CWbounds}. However, as we will discuss below, the latter bounds on the CP-odd operators are more robust than those derived from light fermion EDMs.

\subsection{Triple-gauge-boson interactions} \label{sec:TGCs}

The dimension-six operators defined in~\cref{eq:operators} modify the triple-gauge-boson interaction vertices. We~parameterize these interactions with $V = \gamma, Z$ following the conventions introduced in~\citere{Hagiwara:1986vm}:
\beq \label{eq:LWWV}
\begin{split}
{\cal L}_{WWV} = i g_{WWV}\, \bigg[
& \hspace{0.25mm} g_1^V \left ( W_{\mu \nu}^+ \hspace{0.25mm}W^{- \hspace{0.25mm}\mu}- W^{+ \hspace{0.25mm}\mu}\hspace{0.25mm}W^-_{\mu \nu}\right ) V^\nu \\[2mm]
& + \kappa_V\hspace{0.25mm}W_\mu^+\hspace{0.25mm}W_\nu^-\hspace{0.25mm}V^{\mu \nu} + \frac{\lambda_V}{m_W^2}\, W_{\mu \nu}^+ \hspace{0.25mm}W^{- \hspace{0.25mm}\nu \rho}\hspace{0.25mm}{V_\rho}^\mu \\[2mm]
& + \widetilde{\kappa}_V \hspace{0.25mm}W_\mu^+ \hspace{0.25mm}W_\nu^-\hspace{0.25mm}\widetilde{V}^{\mu \nu} + \frac{\widetilde{\lambda}_V}{m_W^2} \, W_{\mu \nu}^+ \hspace{0.25mm} W^{- \hspace{0.25mm}\nu \rho}\hspace{0.25mm}{\widetilde{V}_\rho}^{\mu} \\[2mm]
& + \widetilde{\xi}_V \hspace{0.5mm} \varepsilon^{\mu \nu \lambda \rho} \hspace{0.25mm} \left(W_{\mu}^{+} \hspace{0.25mm} \partial_{\lambda} W_{\nu}^{-} - W_{\mu}^{-} \partial_{\lambda} \hspace{0.25mm} W_{\nu}^{+}\right) \hspace{0.25mm} V_\rho \hspace{0.25mm} \bigg ] \,.
\end{split}
\eeq
The overall coupling strengths are given by $g_{WW\gamma}= -e$ and $g_{WWZ}= -c_w/s_w \hspace{0.25mm}e$. The field-strength tensors are defined as 
$W_{\mu \nu}^{\pm}= \partial_\mu W_\nu^\pm - \partial_\nu W_\mu^\pm$ and $V_{\mu \nu}= \partial_\mu V_\nu - \partial_\nu V_\mu$, where $W_\mu^\pm$ and $V_\mu$ denote the physical gauge-boson fields. The dual field-strength tensor is defined as $\widetilde{V}_{\mu\nu}= \epsilon_{\mu \nu \lambda \rho}\hspace{0.25mm}V^{\lambda \rho}/2$. Note~that, in contrast to~\citere{Hagiwara:1986vm}, the Lorentz structures proportional to~$g_4^V$~are omitted from~\cref{eq:LWWV} since they are not generated in our case. Additionally, the couplings~$g_5^V$~have been renamed to $\widetilde{\xi}_V$.

By defining $g_1^V = 1 + \Delta g_1^V$ and $\kappa_V = 1 + \Delta \kappa_V$, the shifts $\Delta g_1^V$ and $\Delta \kappa_V$, along with the couplings~$\lambda_V$, $\widetilde{\kappa}_V$, $\widetilde{\lambda}_V$, and $\widetilde{\xi}_V$, can be directly related to the Wilson coefficients of the operators introduced in~\cref{eq:operators}. For the CP-conserving couplings, we obtain the following results:
\beq \label{eq:CPevencoup}
\begin{split}
\Delta g_1^\gamma & = - s_w \hspace{0.25mm} c_w \hspace{0.25mm} \frac{v^2}{\Lambda^2} \hspace{0.5mm} C_{HW\!B} + c_w^2 \hspace{0.5mm} \delta g_1 + s_w^2 \hspace{0.5mm} \delta g_2 \,, \\[2mm]
\Delta g_1^Z & = \frac{s_w^3}{c_w} \frac{v^2}{\Lambda^2} \hspace{0.5mm} C_{HW\!B} - s_w^2 \hspace{0.5mm} \delta g_1 + \left ( 1 + s_w^2 \right ) \delta g_2 \,, \\[2mm]
\Delta \kappa_\gamma & = \frac{c_w^3}{s_w} \frac{v^2}{\Lambda^2} \hspace{0.5mm} C_{HW\!B} + c_w^2 \hspace{0.5mm} \delta g_1 + s_w^2 \hspace{0.5mm} \delta g_2 \,, \\[2mm]
\Delta \kappa_Z & = - s_w \hspace{0.25mm} c_w \hspace{0.25mm} \frac{v^2}{\Lambda^2} \hspace{0.5mm} C_{HW\!B} - s_w^2 \hspace{0.5mm} \delta g_1 + \left ( 1 + s_w^2 \right ) \delta g_2 \,, \\[2mm]
\lambda_\gamma & = \lambda_Z = \frac{3 e}{2 s_w} \frac{v^2}{\Lambda^2} \hspace{0.5mm} C_{W} \,. 
\end{split}
\eeq
A few comments on the above results are in order. First, the relations \cref{eq:CPevencoup} are independent of both the Wilson coefficients $C_{HB}$ and $C_{HW}$. This can be easily understood by noting that, following EWSB, the operators $Q_{HB}$ and $Q_{HW}$ generate $v^2/\Lambda^2$
corrections to the SM gauge-boson kinetic terms. Rescaling these kinetic terms to achieve properly normalized gauge-boson propagators simultaneously eliminates the dependence on $C_{HB}$ and $C_{HW}$ in all triple-gauge-boson vertices. Second, in addition to $C_{HW\!B}$ and $C_W$, the expressions in~\cref{eq:CPevencoup} also depend on the SMEFT-induced shifts $\delta g_1$ and $\delta g_2$ in the $U(1)_Y$ and $SU(2)_L$ gauge couplings, respectively. These shifts account for the SMEFT-induced modifications to the EW input parameters and therefore depend on the chosen EW input scheme within the SMEFT framework. For instance, in the LEP scheme used in~\cref{sec:wbosonmass} --- where $\alpha$, $G_F$, and $m_Z$ are taken as input parameters --- one obtains, after omitting contributions from operators not included in~\cref{eq:operators}:
\beq \label{eq:LEPshifts}
\delta g_1 = -\frac{c_w}{s_w} \frac{v^2}{\Lambda^2} \hspace{0.5mm} C_{HW\!B} \,, \qquad \delta g_2 = 0 \,, \qquad \text{(LEP scheme)} \,. 
\eeq
On the other hand, in the $G_F$ scheme~\cite{Denner:2000bj} used below in~\cref{sec:results}, where $G_F$, $m_Z$, and $m_W$ are chosen as input parameters, the corresponding expressions are given by:
\beq \label{eq:GFshifts}
\delta g_1 = \frac{s_w \hspace{0.25mm} c_w}{c_w^2 - s_w^2} \frac{v^2}{\Lambda^2} \hspace{0.5mm} C_{HW\!B} \,, \qquad \delta g_2 = - \frac{s_w \hspace{0.25mm} c_w}{c_w^2 - s_w^2} \frac{v^2}{\Lambda^2} \hspace{0.5mm} C_{HW\!B} \,, \qquad \text{($G_F$ scheme)} \,. 
\eeq
Third, observe that in the combination $\Delta g_1^V - \Delta \kappa_V$, the contributions from $\delta g_1$ and $\delta g_2$ cancel out. Moreover, EW gauge invariance implies the following relation:
\beq \label{eq:gaugeinvarrel}
\Delta g_1^\gamma - \Delta \kappa_\gamma = - \frac{s_w^2}{c_w^2} \left ( \Delta g_1^Z - \Delta \kappa_Z \right ) \,.
\eeq

In our \recolatwo{} and \POWHEGBOXRES{} implementations of NLO SMEFT effects in diboson production, we adopt the $G_F$ scheme as defined in \cref{eq:GFshifts}. This leads to the following expressions for~$\Delta g_1^V$ and $\Delta \kappa_V$: 
\beq \label{eq:ourEWscheme}
\begin{gathered} 
\Delta g_1^\gamma = 0 \,, \qquad \Delta g_1^Z = -\frac{s_w}{c_w \left ( c_w^2 - s_w^2 \right )} \frac{v^2}{\Lambda^2} \hspace{0.5mm} C_{HW\!B} \,, \\[2mm] 
\Delta \kappa_\gamma = \frac{c_w}{s_w} \frac{v^2}{\Lambda^2} \hspace{0.5mm} C_{HW\!B} \,, \qquad \Delta \kappa_Z = -\frac{2 s_w \hspace{0.25mm} c_w}{c_w^2 - s_w^2} \frac{v^2}{\Lambda^2} \hspace{0.5mm} C_{HW\!B} \,. 
\end{gathered}
\eeq

The CP-odd shifts and couplings do not depend on the adopted EW input scheme. We obtain the following relations:
\beq \label{eq:CPoddcoup} 
\begin{gathered} 
\widetilde{\kappa}_\gamma = -\frac{2 v^2}{\Lambda^2} \hspace{0.5mm} C_{H\widetilde{W}} + \frac{c_w}{s_w} \frac{v^2}{\Lambda^2} \hspace{0.5mm} C_{H\widetilde{W}\!B} \,, \qquad \widetilde{\kappa}_Z = -\frac{2 v^2}{\Lambda^2} \hspace{0.5mm} C_{H\widetilde{W}} - \frac{s_w}{c_w} \frac{v^2}{\Lambda^2} \hspace{0.5mm} C_{H\widetilde{W}\!B} \,, \\[2mm] 
\widetilde{\lambda}_{\gamma} = \widetilde{\lambda}_{Z} = \frac{3 e}{2 s_w} \frac{v^2}{\Lambda^2} \hspace{0.5mm} C_{\widetilde{W}} \,,
\qquad \widetilde{\xi}_\gamma = \widetilde{\xi}_Z = -\frac{2 v^2}{\Lambda^2} \hspace{0.5mm} C_{H\widetilde{W}} \,.
\end{gathered}
\eeq

Before proceeding, we emphasize that~Eqs.~(\ref{eq:CPevencoup}), (\ref{eq:ourEWscheme}), and (\ref{eq:CPoddcoup}) indicate that, at first glance, the anomalous triple-gauge-boson interactions in~\cref{eq:LWWV} are affected by five of the eight dimension-six SMEFT operators introduced in~\cref{eq:operators}. However, the apparent dependence on $C_{H\widetilde{W}}$ in~\cref{eq:CPoddcoup} is actually spurious. To see this, we note that the terms in the Lagrangian~\cref{eq:LWWV} involving the Wilson coefficient $C_{H\widetilde{W}}$ can be expressed in momentum space as:
\beq \label{eq:spurious}
-i g_{WWV} \hspace{0.5mm} \frac{2 v^2}{\Lambda^2} \hspace{0.5mm} C_{H\widetilde{W}} \hspace{0.5mm} \varepsilon_{\mu \nu \lambda \rho} \hspace{0.25mm} \left ( p_{V}^{\rho} + p_{W^{+}}^{\rho} + p_{W^{-}}^{\rho} \right ) \,. 
\eeq
For simplicity, the polarization vectors of the external gauge bosons have been omitted here, and all gauge-boson four-momenta are taken as incoming. Consequently, four-momentum conservation ensures that~\cref{eq:spurious} vanishes algebraically, implying that the triple-gauge-boson interactions~\cref{eq:LWWV} effectively depend on only four of the eight dimension-six SMEFT operators introduced in \cref{eq:operators}. Note that the non-contribution of the operator $Q_{H\widetilde{W}}$ to the Lagrangian ${\cal L}_{WWV}$ can also be understood from the fact that it can be written as a total derivative after EWSB. The results presented in this subsection were obtained using the \noun{SMEFTsim 3.0} package~\cite{Brivio:2020onw} and independently verified by hand. Selected results were also cross-checked using~\noun{SmeftFR}~v3~\cite{Dedes:2023zws}.

\subsection{Discussion} \label{sec:discussion}

We are now ready to specify representative benchmark scenarios for the set of Wilson coefficients associated with the operators in~\cref{eq:operators}. It is worth emphasizing that~\cref{eq:LWWV} does not depend on the Wilson coefficients $C_{HB}$ and $C_{HW}$. On the other hand, the Wilson coefficient $C_{HW\!B}$ influences the triple-gauge-boson interactions and is generally subject to strong constraints from both the measured $W$-boson mass and the decay process $h \to \gamma \gamma$. Taking this into account, we adopt the following benchmark values for the Wilson coefficients of the CP-even dimension-six operators:
\beq \label{eq:CPevenbench} 
\begin{gathered} 
\frac{C_{HB}}{\Lambda^2} = 4.0 \cdot 10^{-2}\, {\rm TeV}^{-2} \,, 
\qquad \frac{C_{HW}}{\Lambda^2} = -1.2 \cdot 10^{-1}\, {\rm TeV}^{-2} \,, \\[2mm] 
\frac{C_{HW\!B}}{\Lambda^2} = 1.0 \cdot 10^{-2}\, {\rm TeV}^{-2} \,,
\qquad \frac{C_W}{\Lambda^2} = 1 \, {\rm TeV}^{-2}\,.  
\end{gathered}
\eeq
Here, the values of $C_{HB}$, $C_{HW}$, and $C_{HW\!B}$ are chosen such that they induce a shift in the $W$-boson mass that remains experimentally viable, while keeping the Higgs signal strengths for $h \to WW$, $h \to ZZ$, and $h \to \gamma \gamma$ close to their SM expectations. Simultaneously, the $h \to \gamma Z$ decay rate is enhanced by approximately a factor of two compared to the SM. Notably, this pattern of deviations in the Higgs decay rates is supported by LHC~Run~2~data~\cite{ATLAS:2020qdt,CMS:2020gsy,ATLAS:2023yqk}. The chosen value of $C_W$, on the other hand, aligns with that used in the recent analysis~\cite{ElFaham:2024uop}.

As discussed previously, the nominally strongest constraints on the CP-violating Wilson coefficients originate from the light-fermion EDMs, particularly the electron one. In this context, it is important to note that every term in~\cref{eq:deovere} is proportional to the SM electron Yukawa coupling~$y_e$. Consequently, if $y_e$ is significantly suppressed or vanishes due to new-physics effects, the bounds from $d_e/e$ become much less stringent. Within~the SMEFT framework, such a cancellation can be readily realized by including the Yukawa-like operator
\beq \label{eq:eYukawa}
Q_{eH}= \left (L H e \right ) H^\dagger H \,,
\eeq
and appropriately tuning the corresponding Wilson coefficient such that $y_e - v^2/\Lambda^2 \hspace{0.5mm}C_{eH}\simeq 0$. While~such a scenario requires considerable fine-tuning, it is phenomenologically permissible, as current experimental limits allow for a vanishing electron Yukawa coupling without conflict~\cite{Altmannshofer:2015qra,ATLAS:2019old}. A~comparable argument holds for the neutron EDM, where light-quark contributions can similarly be suppressed using the same mechanism. As a result, the associated constraints are found to be rather weak~\cite{Haisch:2019xyi}. This demonstrates that, in general, the EDM constraints are more sensitive to SMEFT deformations than the bounds on CP-violating operators that arise, for instance, from the $h \to \gamma \gamma$ decay. A similar observation holds for the limits on anomalous triple-gauge-boson couplings derived from flavor physics~\cite{Bobeth:2015zqa}, which can also be relaxed by introducing new sources of flavor violation within the SMEFT. Hence, in the following, we will not consider any constraints on the operators in~\cref{eq:operators} that may arise from low-energy measurements.

Since the CP-odd operators do not enter~\cref{eq:dMWoMWtheo} and contribute to~\cref{eq:digammasignalstrength} in the same form as their CP-even counterparts, the constraints from the $W$-boson mass and the $h \to \gamma \gamma$ decay on the CP-violating gauge-Higgs operators can effectively be avoided by selecting, for example:
\beq \label{eq:CPoddbench} 
\begin{gathered}
\frac{C_{H\widetilde{B}}}{\Lambda^2} = 4.0 \cdot 10^{-2}\, {\rm TeV}^{-2} \,,
\qquad \frac{C_{H\widetilde{W}}}{\Lambda^2} = -1.2 \cdot 10^{-1}\, {\rm TeV}^{-2} \,, \\[2mm] 
\frac{C_{H\widetilde{W}\!B}}{\Lambda^2} = 1.0 \cdot 10^{-2}\, {\rm TeV}^{-2} \,,
\qquad \frac{C_{\widetilde{W}}}{\Lambda^2} = 1 \, {\rm TeV}^{-2}\,.
\end{gathered}
\eeq
With regard to the choice of $C_{\widetilde{W}}$, we note that the same value was adopted in~\citere{ElFaham:2024uop}, allowing for a direct comparison between their results and ours.

\section{Phenomenological analysis} \label{sec:results}

Among the various diboson production channels, the $W^\pm Z$ process stands out as particularly well-suited for polarization studies, as demonstrated by the recent measurements of polarization fractions by the ATLAS and CMS collaborations using LHC Run 2 data~\cite{ATLAS:2019bsc,CMS:2021icx,ATLAS:2022oge,ATLAS:2024qbd}. Besides its relatively high production rate, the three-lepton decay channel provides excellent signal purity and enables the reconstruction of the final state via single-neutrino reconstruction. In this section, we present NLO+PS accurate predictions for $W^+ Z$ production with leptonic decays, based on the SMEFT benchmark scenarios involving the dimension-six operators introduced in~\cref{eq:operators}.

\subsection{Numerical input} \label{sec:input}

All SM input parameters are taken from the most recent PDG review~\cite{ParticleDataGroup:2024cfk}. For the on-shell masses and the total decay widths of the EW gauge bosons, we use the following values:
\beq \label{eq:input1}
m_W^{\rm OS}= 80.369 \, {\rm GeV}\,, \quad m_Z^{\rm OS}= 91.188 \, {\rm GeV}\,, \quad \Gamma^{\rm OS}_W = 2.085 \, {\rm GeV}\,, \quad \Gamma^{\rm OS}_Z = 2.4955 \, {\rm GeV}\,. 
\eeq
The on-shell masses are converted to pole masses via the relation~\cite{Bardin:1988xt}
\beq \label{eq:OStopole}
m_V = \frac{m_V^{\rm OS}}{\sqrt{1 + \left ( \displaystyle \frac{\Gamma^{\rm OS}_V}{m_V^{{\rm OS}}}\right )^2}}\,, \qquad
\Gamma_V = \frac{\Gamma_V^{\rm OS}}{\sqrt{1 + \left ( \displaystyle \frac{\Gamma^{\rm OS}_V}{m_V^{{\rm OS}}}\right )^2}}\,,
\eeq
with $V = W, Z$, yielding
\beq \label{eq:polemasses}
m_W = 80.342 \, {\rm GeV}\,, \qquad m_Z = 91.154 \, {\rm GeV}\,,\qquad
\Gamma_W = 2.084 \, {\rm GeV}\,, \qquad \Gamma_Z = 2.4946 \, {\rm GeV}\,. 
\eeq
Within the $G_F$ scheme, the EW input is defined by the pole masses of weak bosons together with the Fermi constant:
\beq \label{eq:Fermiconstant}
G_F = 1.1663788 \cdot 10^{-5}\, {\rm GeV}^{-2}\,.
\eeq

For off-shell calculations, we make use of the complex-mass scheme~\cite{Denner:2005fg,Denner:2019vbn,Denner:2006ic}. Consequently, the electromagnetic coupling constant and the sine squared of the weak mixing angle are determined as follows:
\beq \label{eq:alphasine}
\begin{gathered} 
\alpha = \frac{\sqrt{2}\hspace{0.125mm}G_F}{\pi}\, m_W^2 \left ( 1 - \frac{m_W^2}{m_Z^2}\right ) = \frac{1}{132.222}\,, \\[2mm] 
s_w^2 = 1 - \frac{\mu_W^2}{\mu_Z^2}= 0.2232 - 0.0011 \hspace{0.25mm}i \,, 
\end{gathered}
\eeq
where the complex squared masses $\mu_W^2$ and $\mu_Z^2$ of the gauge bosons are defined in~\cref{eq:muV2}. It~is~important to note that the complex-valued definition of $s_w^2$ guarantees gauge invariance and ensures a consistent treatment of unstable particles at the loop level. However, when evaluating the shifts~$\Delta g_1^V$ and $\Delta \kappa_V$, as well as the couplings $\lambda_V$, $\widetilde{\kappa}_V$, $\widetilde{\lambda}_V$, and $\widetilde{\xi}_V$ defined in~\cref{eq:CPevencoup,eq:CPoddcoup}, we explicitly use the real parts of the sine and cosine of the weak mixing angle. This is motivated by the fact that the~$s_w$ and~$c_w$ terms in the Lagrangian~\cref{eq:LWWV} arise from the tree-level, real-valued rotation of the weak eigenstate fields in the operators~\cref{eq:operators} into the mass-eigenstate fields that enter the triple-gauge-boson interactions. This approach ensures that the small imaginary component of~$s_w^2$ in the complex-mass scheme does not induce mixing between CP-even and -odd interactions. In~the DPA, all EW gauge boson widths are neglected except for the two resonant $s$-channel bosons, ensuring that the weak mixing angle and couplings are purely real.

In our MC simulations, all leptons are treated as massless, and the five-flavor scheme is employed. Quark mixing between generations is neglected. The parton distribution functions~(PDFs) and the strong coupling constant $\alpha_s$ are evaluated using the \noun{Lhapdf} interface~\cite{Buckley:2014ana}, with the \noun{Nnpdf31\_nlo\_as\_0118} set~\cite{NNPDF:2017mvq}, corresponding to $\alpha_s(m_Z) = 0.118$. The renormalization and factorization scales are set to the arithmetic mean of the pole masses of the bosons involved in the process. For $W^\pm Z$ production, this means:
\beq \label{eq:muchoice}
\mu_R = \mu_F = \frac{m_W + m_Z}2 = 85.748 \, {\rm GeV}\,.
\eeq

\subsection{Selection cuts} \label{sec:setup}

The signal process analyzed in this section is $pp \to W^+ Z \to e^+\nu_e \hspace{0.5mm}\mu^+ \mu^-$ at a CM energy of $\sqrt{s}= 13 \, {\rm TeV}$. Details on the event generation using the \POWHEGBOXRES{} implementation are provided in Appendix~\ref{app:MC}. Throughout this study, leptons are treated as dressed by including photon radiation within a cone of resolution radius $R =0.1$. As the analysis includes only NLO QCD corrections, QED radiation is accounted for solely via the PS. Parton showering is performed using \noun{Pythia~8.2}~\cite{Sjostrand:2014zea}, which incorporates QCD and QED shower effects, hadronization, and multi-parton interactions (MPI). 

We examine two experimental setups. The inclusive setup is defined by an invariant-mass cut imposed on same-flavor, opposite-sign lepton pairs originating from the $Z$-boson decays. In our analysis, we employ:
\beq \label{eq:inclusive}
81 \, {\rm GeV}< m_{\mu^+\mu^-}< 101 \, {\rm GeV}\,.
\eeq
For the fiducial setup, we adopt the selection criteria used in the recent ATLAS analysis~\cite{ATLAS:2022oge}, which besides~\cref{eq:inclusive} include the following selection requirements:
\beq \label{eq:fiducial}
\begin{gathered}
p_{T, e}> 20 \, {\rm GeV}\,, \qquad p_{T, \mu}> 15 \, {\rm GeV}\,, \qquad |\eta_{l}| < 2.5 \,, \\[2mm]
\Delta R_{e\mu}> 0.3 \,, \qquad \Delta R_{\mu^+ \mu^-}> 0.2 \,, \qquad m_{T,W}> 30 \, {\rm GeV}\,.
\end{gathered}
\eeq
Here, $p_{T,l}$ refers to the transverse momentum of the relevant charged lepton, and $\eta_l$ denotes its pseudorapidity. The variables $\Delta R_{e\mu}$ and $\Delta R_{\mu^+ \mu^-}$ represent the angular separations between the corresponding lepton pairs. The transverse mass of the $W$ boson is given by
\beq 
m_{T,W}= \sqrt{2 \hspace{0.25mm}p_{T, e} \hspace{0.25mm} p_{T, \rm miss} \left ( 1 - \cos \Delta \phi_{e \hspace{0.25mm} \rm miss} \right )}\,, 
\eeq
where $p_{T, \rm miss}$ is the missing transverse momentum --- serving as a proxy for the neutrino --- and $\Delta \phi_{e \hspace{0.25mm}\rm miss}$ represents the azimuthal angle between the electron and the missing transverse momentum~vector.

\subsection{Off-shell results} \label{sec:validation}

{\renewcommand{\arraystretch}{1.5}
\begin{table}[!t]
\begin{center}
\begin{tabular}{|c|c|c|}
\hline contribution & this work & \citere{ElFaham:2024uop}\\ 
\hline \hline SM & \phantom{+}$ 35.30(3) \, {\rm fb}$ & \phantom{+}$ 35.35(1) \, {\rm fb}$ \\
\hline $Q_W$ (lin.) & $-0.996(3) \, {\rm fb}$ & $-0.997(2) \, {\rm fb}$ \\
$Q_W$ (quad.) & \phantom{+}$ 6.57(1) \, {\rm fb}$ & \phantom{+}$ 6.58(1) \, {\rm fb}$ \\
\hline $Q_{\widetilde{W}}$ (lin.) & $-0.062(2) \, {\rm fb}$ & $-0.059(1) \, {\rm fb}$ \\
$Q_{\widetilde{W}}$ (quad.) & \phantom{+}$ 6.72(1) \, {\rm fb}$ & \phantom{+}$ 6.71(1) \, {\rm fb}$ \\
\hline
\end{tabular}\qquad
\end{center}
\vspace{0mm}
\caption{Fiducial NLO QCD cross sections for the off-shell $W^+ Z$ production process at $\sqrt{s}= 13 \, {\rm TeV}$ within the ATLAS fiducial phase space, as defined in \cref{eq:fiducial}. Results obtained using \recolatwo{} and \POWHEGBOXRES{} are compared with those from~\citere{ElFaham:2024uop}, which employed \SMEFTNLO{} and \MadGraph. In the SMEFT cases, the label ``lin.'' (``quad.'') refers to the size of the linear (quadratic) BSM contribution of order $1/\Lambda^2$~($1/\Lambda^4$). The values $C_W/\Lambda^2 = 1 \, {\rm TeV}^{-2}$ or $C_{\widetilde{W}}/\Lambda^2 = 1 \, {\rm TeV}^{-2}$ are employed for the Wilson coefficients. The uncertainties displayed in parentheses correspond to MC-integration errors. \label{tab:comp}}
\end{table}
}

\begin{figure}[t!]
\centering
\includegraphics[width=.49\textwidth,page=1]{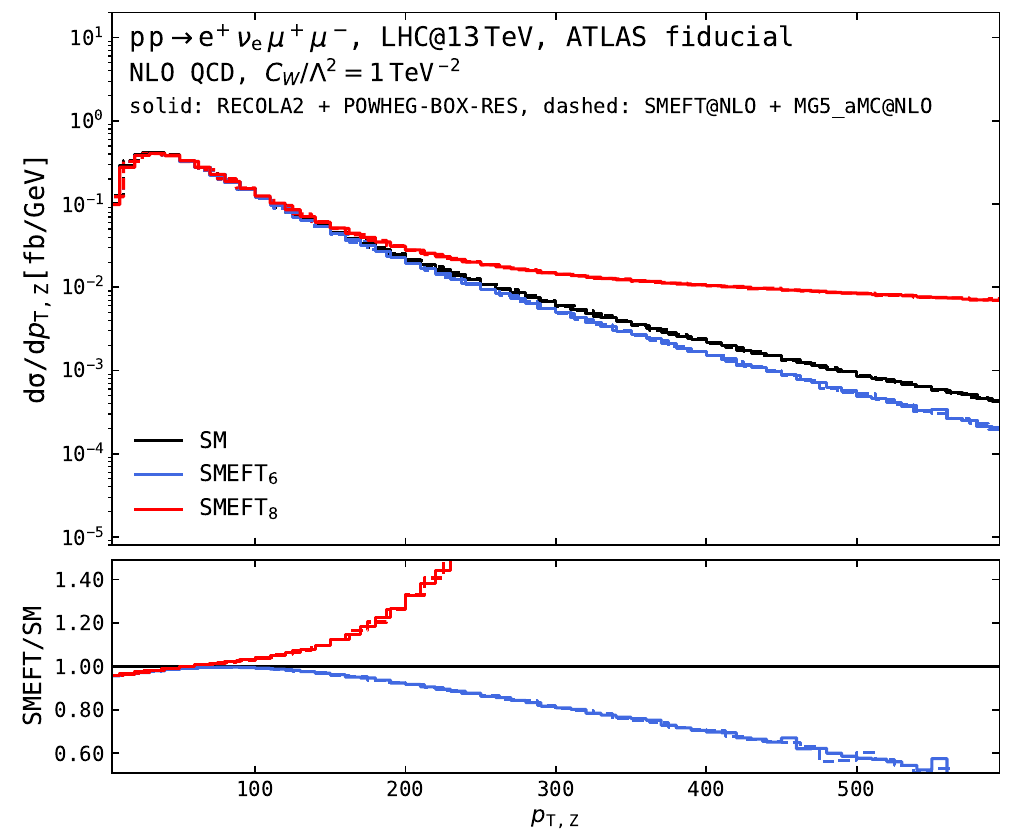}
\includegraphics[width=.49\textwidth,page=1]{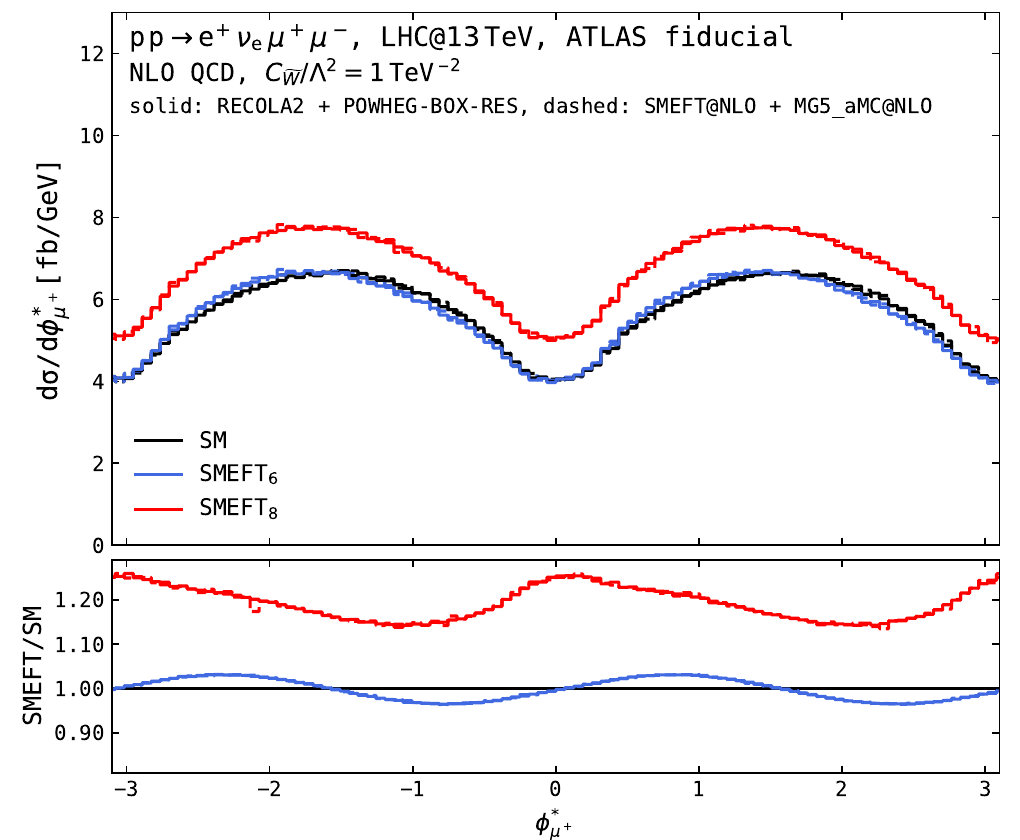}
\vspace{2mm}
\caption{Differential NLO QCD distributions in the $Z$-boson transverse momentum~($p_{T, Z}$) in the right panel, and in the antimuon decay angle in the $Z$-boson rest frame~($\phi_{\mu^+}^\ast$) in the left panel, within the ATLAS fiducial phase space, as defined in \cref{eq:fiducial}. A~comparison is made between the $\sqrt{s}= 13 \, {\rm TeV}$ results obtained with \recolatwo{} and \POWHEGBOXRES{}~(solid lines) and those using \SMEFTNLO{} and \MadGraph{}~(dashed lines), as presented in~\citere{ElFaham:2024uop}. In the left~(right) panel the displayed BSM results correspond to $C_W/\Lambda^2 = 1 \, {\rm TeV}^{-2}$ ($C_{\widetilde{W}}/\Lambda^2 = 1 \, {\rm TeV}^{-2}$). The~notation~$\text{SMEFT}_6$~($\text{SMEFT}_8$) refers to the sum of the SM and SMEFT contributions, including terms up to order $1/\Lambda^2$ ($1/\Lambda^4$). The lower sections of the panels show the ratios between the BSM and SM predictions. \label{fig:comp}}
\end{figure}

As mentioned earlier in~\cref{sec:recola}, our new implementation of NLO SMEFT effects in diboson production using \recolatwo{} and \POWHEGBOXRES{} has been validated by comparison with the results of~\citere{ElFaham:2024uop}, which utilized \SMEFTNLO{} in conjunction with \MadGraph{}. Table~\ref{tab:comp}~presents the fiducial NLO~QCD cross sections for full off-shell $W^+ Z$ production at $\sqrt{s}= 13 \, {\rm TeV}$, evaluated within the ATLAS fiducial phase space defined in \cref{eq:fiducial}. We compare the SM predictions as well as the BSM contributions linear and quadratic in the Wilson coefficients, employing $C_W/\Lambda^2 = 1 \, {\rm TeV}^{-2}$ and $C_{\widetilde{W}}/\Lambda^2 = 1 \, {\rm TeV}^{-2}$. Within the statistical uncertainties from the MC integration, we observe excellent agreement between the two approaches. The consistency between the two setups has also been verified at the differential level. \cref{fig:comp} presents the validation results for the transverse momentum of the $Z$~boson~($p_{T, Z}$) and the distribution of the antimuon decay angle in the $Z$-boson rest frame~($\phi_{\mu^+}^\ast$). Once again, very good agreement is seen for both the SM and all displayed BSM distributions. It is important to note that the differences between the results from the two setups remain within the MC-integration errors in each bin, not only in the bulk of the distributions but also in regions where the differential cross section is suppressed, such as the high-$p_{T,Z}$ tail illustrated in the left panel. 

\begin{figure}[t!]
\centering
\includegraphics[width=.49\textwidth,page=1]{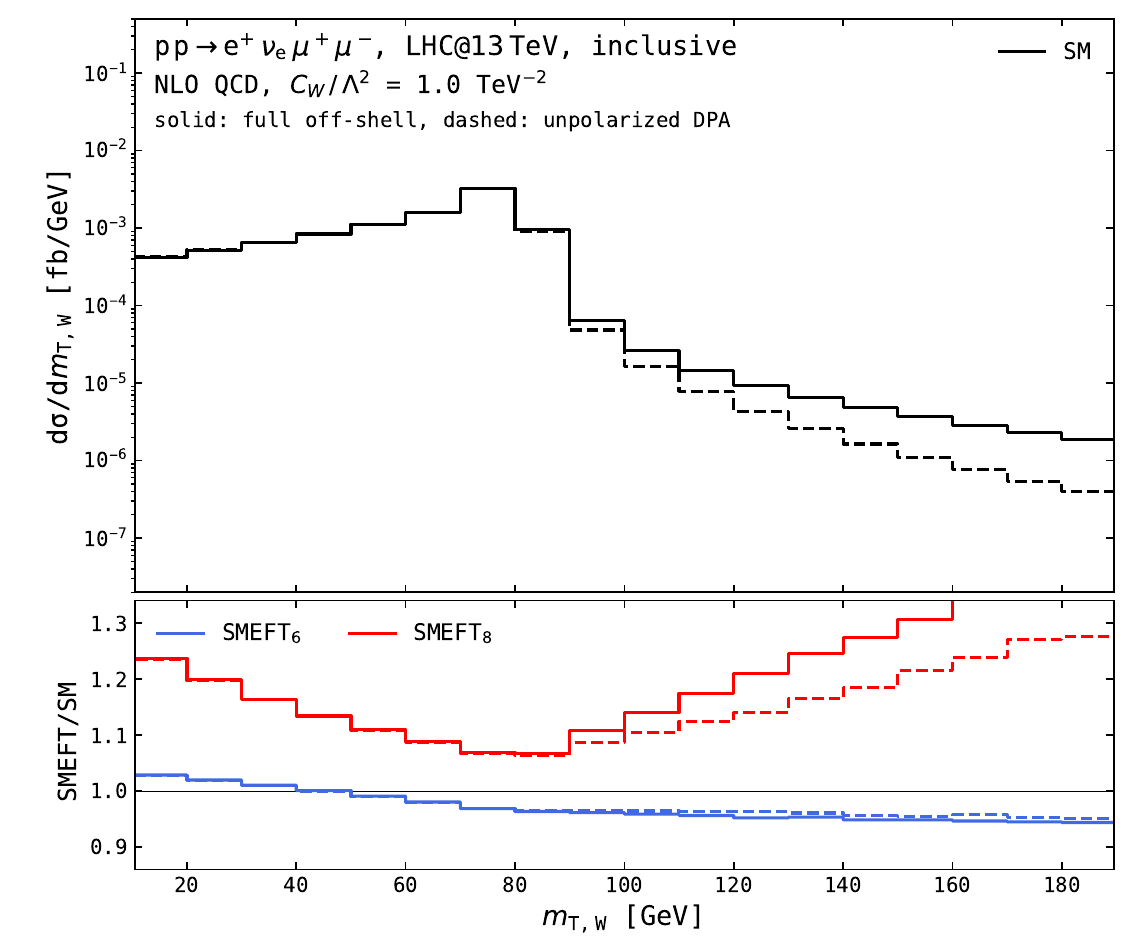}
\includegraphics[width=.49\textwidth,page=2]{off_on_shell.pdf}
\vspace{2mm}
\caption{Transverse-mass distributions of the $W$ boson ($m_{T,W}$) at $\sqrt{s}= 13 \, {\rm TeV}$ are shown for both the inclusive (left panel) and fiducial (right panel) setups, as defined in \cref{eq:inclusive,eq:fiducial}, respectively. Solid lines represent the full off-shell predictions, while dashed lines correspond to the unpolarized DPA results. Each panel is divided into two sections, with the absolute SM prediction in the upper section and ratios of BSM to SM results in the lower section within the same approximation. The BSM curves employ $C_W/\Lambda^2 = 1 \, {\rm TeV}^{-2}$, with $\text{SMEFT}_6$ and $\text{SMEFT}_8$ defined as in~\cref{fig:comp}. \label{fig:mtwoff}}
\end{figure}

\begin{figure}[t!]
\vspace{4mm}
\centering
\includegraphics[width=.65\textwidth]{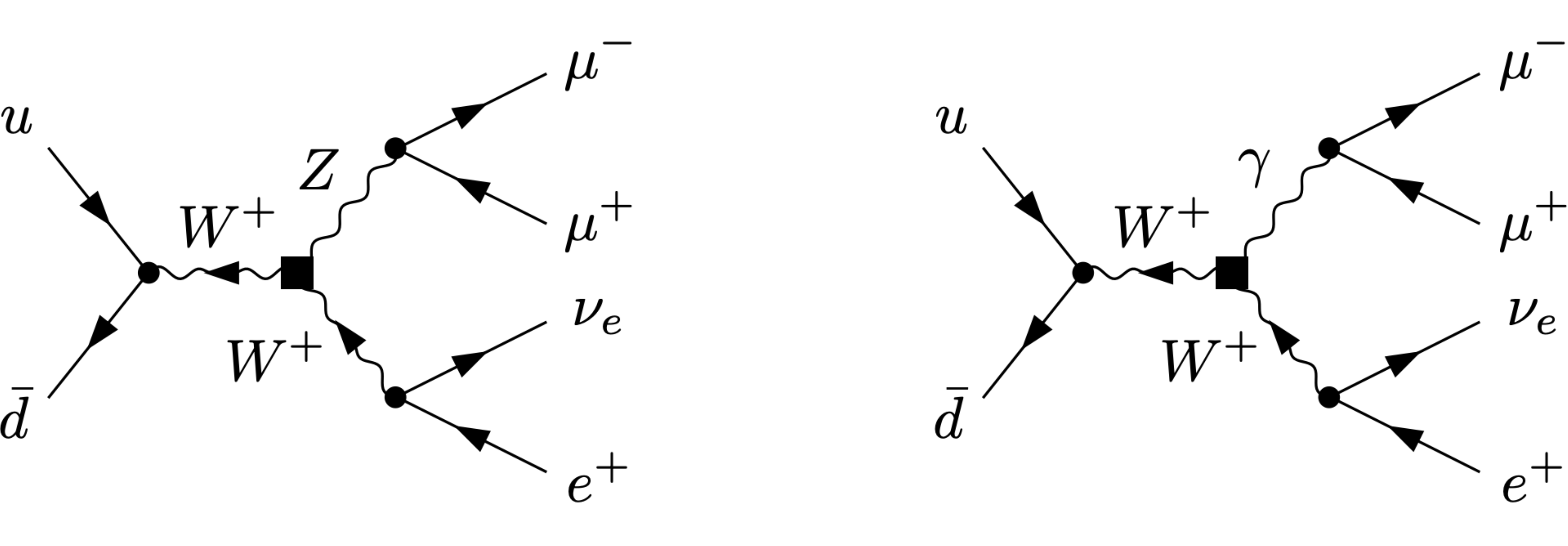}
\vspace{0mm}
\caption{Contributions from the dimension-six SMEFT operator $Q_W$ to the triple-gauge-boson couplings $WWZ$~(left) and~$WW\gamma$~(right) in $W^+ Z$ production with leptonic decays. The black boxes represent the operator insertions. Both diagrams are included in the full off-shell calculation, whereas only the left diagram is taken into account in the DPA. \label{fig:WZ_eft_diag}}
\end{figure}

The off-shell characterization of $W^\pm Z$ production at the LHC has been extensively explored in the literature~\cite{Baglio:2017bfe,Azatov:2017kzw,Panico:2017frx,Franceschini:2017xkh,Chiesa:2018lcs,Liu:2018pkg,Grojean:2018dqj,Baglio:2018bkm,Azatov:2019xxn,Baglio:2019uty,Baglio:2020oqu,Ellis:2020unq,Degrande:2021zpv,Degrande:2023iob,Aoude:2023hxv,Degrande:2024bmd,ElFaham:2024uop,Banerjee:2024eyo,Thomas:2024dwd}. A leading-order~(LO) comparison with the NWA was performed in~\citere{ElFaham:2024uop}, providing a qualitative insight into off-shell effects. The \POWHEGBOXRES{} implementation introduced in this work facilitates a comprehensive and quantitative study of genuine off-shell effects at NLO QCD accuracy, through a comparison between full off-shell predictions and those using the unpolarized DPA. In~\cref{fig:mtwoff}, we consider the spectrum of the $W$-boson transverse mass~($m_{T,W}$), which serves as a sensitive observable for distinguishing between the two approaches. Our results for the inclusive and fiducial experimental setups, defined in \cref{eq:inclusive,eq:fiducial}, are displayed in the left and right panels, respectively. It~is evident from the plots that below the $W$-boson mass threshold, where $m_{T,W}< m_W$, the DPA provides an excellent approximation to the full off-shell results in the SM as well as in the studied BSM scenario with $C_W/\Lambda^2 = 1 \, {\rm TeV}^{-2}$. Above the $W$-boson mass threshold, the discrepancy between the two methods increases with $m_{T,W}$. This behavior arises because the DPA systematically underestimates the differential cross section by neglecting singly-resonant and non-resonant contributions, which become increasingly important in regions where $m_{T,W}> m_W$. The discrepancy becomes especially evident when SMEFT contributions are included. In particular, the DPA yields a more accurate approximation for the SM and $\text{SMEFT}_6$ case compared to $\text{SMEFT}_8$. In the~SM, the region of large~$m_{T,W}$, though suppressed, receives notable contributions from diagrams not captured by the DPA, shown on the right side of~\cref{fig:WZLO}. Despite this, $t$-channel doubly-resonant diagrams continue to provide a sizable contribution. Consequently, the full off-shell prediction exceeds the DPA estimate by about $50\%$ at $m_{T,W} \simeq 100 \, {\rm GeV}$. Owing to interference suppression in the dominant helicity configurations~\cite{Azatov:2017kzw}, the behavior of the $\text{SMEFT}_6$ contributions closely resembles that of the~SM. For the squared SMEFT terms, only two diagrams contribute, both involving an~$s$-channel $W$~boson. These diagrams differ by the nature of the second $s$-channel propagator: one includes a $Z$ boson, the other an off-shell photon. The $Q_W$ operator alters the triple-gauge-boson couplings ---~specifically the $WWZ$ and $WW\gamma$ vertices~--- by introducing a new Lorentz structure, as shown in~\cref{eq:LWWV,eq:CPevencoup}. The corresponding Feynman diagrams are shown in~\cref{fig:WZ_eft_diag}. This modification leads to an amplitude that grows with energy, a~hallmark behavior in effective field theories, in both diagrams. In contrast, the DPA captures only the partial off-shell contributions from the diagram involving the $Z$ boson, while neglecting the photon-mediated diagram. This provides a qualitative explanation for the discrepancies seen in the~$\text{SMEFT}_8$ case. Finally, we observe that the differences between the full off-shell and unpolarized DPA predictions are sensitive to the selection cuts applied in the experimental analysis, with slightly larger deviations observed in the fiducial phase space compared to the inclusive case. 

A comparison between the full off-shell calculation and the DPA with unpolarized bosons has also been carried out at the integrated level, considering both inclusive and fiducial configurations. In addition to the SM, this analysis includes the effects of the $Q_W$ and $Q_{\widetilde{W}}$ operators. The corresponding cross sections are presented in~\cref{tab:IntegOffvsDPA}. Due to the effects discussed in \cref{fig:mtwoff}, the DPA reproduces the off-shell modeling of the linear contribution from $Q_W$ to within $1\%$. However, when the quadratic term is included, the discrepancy increases to about $4\%$. The CP-violating operator~$Q_{\widetilde{W}}$ exhibits similar behavior at the quadratic level but shows a notably different pattern at the linear level. In both experimental configurations analyzed, off-shell effects, complete spin correlations, and the presence of real QCD radiation contribute to a non-vanishing interference between the CP-even SM amplitude and the CP-odd SMEFT amplitude. In contrast, if both bosons were strictly on-shell, this interference would be exactly zero at LO in QCD~\cite{ElFaham:2024uop}. This explains why, in the DPA --- which retains partial off-shell effects and spin correlations between production and decay --- the linear term is strongly suppressed compared to the full off-shell calculation. This suppression is no longer present in the fiducial setup, where kinematic cuts obstruct the cancellation of interference terms between different polarization states.

{\renewcommand{\arraystretch}{1.5}
\begin{table}[!t]
\begin{center}
\begin{tabular}{|c|cc|cc|}
\hline
& \multicolumn{2}{c|}{$\sigma$ [fb], inclusive, \cref{eq:inclusive}}& \multicolumn{2}{c|}{$\sigma$ [fb], ATLAS fiducial, \cref{eq:fiducial}}\\
\hline
 & full off-shell & DPA unpolarized & full off-shell & DPA unpolarized \\
\hline
 SM & $ 98.30(1) $ & $ 97.28(1)$ & $ 35.29(1) $ & $ 34.64(1) $\\
\hline
$Q_W$ (lin.) & $ -1.392(1) $ & $ -1.388(1) $ & $ -0.9835(8) $ & $ -0.9748(7) $\\ 
$Q_W$ (quad.) & $ 12.430(2) $ & $ 12.077(9) $ & $ 6.563(2) $ & $ 6.299(1) $\\
\hline
 $Q_{\widetilde{W}}$ (lin.) & $ 0.194(1) $ & $ 0.007(1) $ & $ 0.0644(9) $ & $ -0.0130(7) $\\
$Q_{\widetilde{W}}$ (quad.) & $ 12.787(2)$ & $ 12.419(1) $ & $ 6.721(2) $ & $ 6.443(1) $\\
\hline
\end{tabular}\qquad
\end{center}
\vspace{0mm}
\caption{Inclusive and fiducial NLO QCD cross sections in units of $\rm fb$ for $W^+ Z$ production at $\sqrt{s} = 13 \, {\rm TeV}$. The results of the full off-shell calculation are compared to the results assuming unpolarized bosons within the DPA. The labelling and notation follow the conventions established in~\cref{tab:comp}. The reported BSM cross sections correspond to a coupling value of $C_W/\Lambda^2 = 1 \, {\rm TeV}^{-2}$ and $C_{\widetilde{W}}/\Lambda^2 = 1 \, {\rm TeV}^{-2}$. \label{tab:IntegOffvsDPA}}
\end{table}
}

Having validated our new \POWHEGBOXRES{} code for off-shell fiducial and differential cross sections and examined the differences between full off-shell and unpolarized DPA predictions, we now turn our attention to polarized signatures. 

\subsection{Singly-polarized signals} \label{sec:wpol}

\begin{figure}[t!]
\centering
\includegraphics[width=.99\textwidth,page=1]{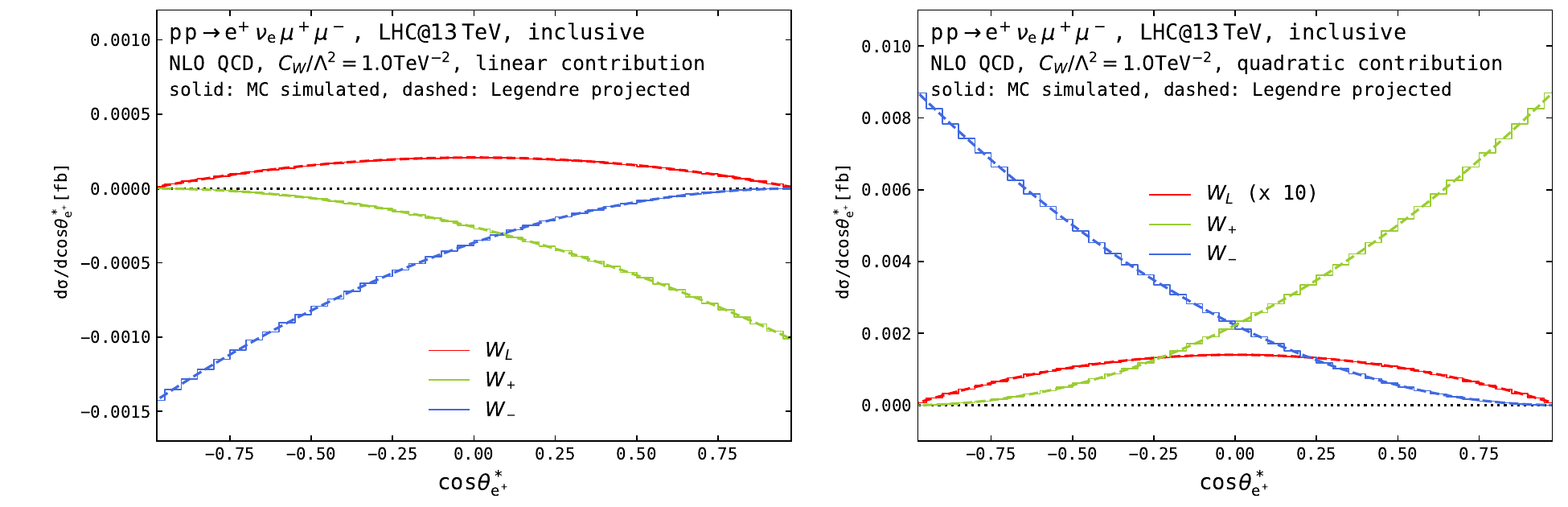}
\vspace{2mm}
\caption{Distributions of the cosine of the polar decay angle of the positron in the $W$-boson rest frame ($\cos\theta^\ast_{\Pe^+}$) at $\sqrt{s} = 13 \, {\rm TeV}$ are shown for the inclusive setup defined in~\cref{eq:inclusive}. The~analysis considers a $W$~boson with a fixed polarization state ($L,+,-$) and an unpolarized $Z$ boson. Solid~lines represent the results of a direct simulation of singly-polarized signals in the DPA, while dashed lines are obtained by projecting the unpolarized DPA results onto Legendre polynomials of degree up to~2. The left and right figures show the linear and quadratic SMEFT contributions, respectively, corresponding to the choice~$C_W/\Lambda^2 = 1 \, {\rm TeV}^{-2}$. Please note that the $W_L$ contribution in the quadratic case, shown in the right plot, has been inflated by a factor of $10$.}
 \label{fig:WZinc_Legendre}
 \end{figure}

The validation of the polarization implementation in the case of the SM, already carried out in the context of \citere{Pelliccioli:2023zpd}, has been performed successfully for our new MC~code as well. To verify~the correctness of the polarization selection in the case of SMEFT contributions,~\cref{fig:WZinc_Legendre} presents the distributions of the cosine of the polar decay angle of the positron in the $W$-boson rest frame~($\cos \theta^\ast_{\Pe^+}$). We~consider $W^+ Z$ production involving a $W$ boson in a fixed polarization state ($L,+,-$) and an unpolarized $Z$~boson in the inclusive setup described in \cref{eq:inclusive}. Solid curves correspond to direct simulations of singly-polarized signals using the DPA, while dashed curves result from projecting the unpolarized DPA output onto Legendre polynomials (see, for instance, \citere{Ballestrero:2017bxn} for details). The left (right) figure displays the linear (quadratic) SMEFT contributions. The BSM curves correspond to the choice $C_W/\Lambda^2 = 1 \, {\rm TeV}^{-2}$. An analogous validation of the $Z$-boson polarizations was conducted by examining the antimuon decay angle distributions, revealing good agreement with the direct MC simulations involving a polarized $Z$ boson produced in association with an unpolarized $W$~boson. For all presented results, there is excellent consistency between the direct MC simulation and the Legendre-projection method. The~observed agreement provides a non-trivial validation of the modified \recolatwo{} interface. The polarization projection was also successfully validated at the amplitude level using selected phase-space points for both Born and real-emission kinematics, although the corresponding results are not shown here. 

As shown in full off-shell calculations~\cite{Azatov:2017kzw,Panico:2017frx,Franceschini:2017xkh,Azatov:2019xxn,Baglio:2019uty,Degrande:2024bmd,ElFaham:2024uop}, including QCD corrections “resurrects” the interference between the SM and dimension-six operators even for definite helicities of the intermediate gauge bosons within the DPA --- at a level of about $2\%$ of the SM cross~section. As~can already be seen in the left panel of~\cref{fig:WZinc_Legendre}, the insertion of the $Q_W$ operator leads to a positive (negative) linear dimension-six contribution for longitudinal (transverse) $W$ bosons. Due to energy suppression, the squared term for the longitudinal mode is smaller than the interference term, whereas the transverse mode, featuring an energy-growing behavior, shows the opposite trend~\cite{Azatov:2017kzw}. In the SM, the longitudinal polarization accounts for roughly $20\%$ of the total (unpolarized) cross section. In contrast, its contribution to the squared SMEFT terms is below $2\%$, indicating that the $Q_W$ and $Q_{\widetilde{W}}$ operators predominantly affect the transverse modes~\cite{Azatov:2017kzw,Baglio:2019uty}. \cref{sec:wzpol} will clarify that a $Q_W$ operator insertion can only produce a longitudinal $W$ ($Z$) boson in association with a transverse $Z$ ($W$) boson, a consequence of the Lorentz structure introduced by $Q_W$ in the $WWZ$ vertex. Although the CP-odd linear contribution is suppressed due to the CP-even nature of the SM amplitude, it is non-vanishing and exhibits opposite signs for the two polarization states. As~also seen for full off-shell $W^\pm Z$ production~\cite{ElFaham:2024uop}, this CP-odd interference would vanish in a $2 \to 2$ process~\cite{Helset:2017mlf,Azatov:2016sqh}, but persists in the $2 \to 4$ case due to spin correlations between production and decay stages, as well as the impact of selection cuts. Since in~Table~\ref{tab:Wpol} an unpolarized $Z$ boson is understood, the spin correlations related to the $Z$ boson are retained in the DPA modeling. In~the transverse polarization case, partial spin correlations associated with the $W$ boson are taken into account, due to the coherent superposition of its right- and left-handed helicity states. Notably, the interference between longitudinal and transverse modes --- estimated by subtracting the individual longitudinal and transverse contributions from the total unpolarized cross section --- is found to be consistent with zero in the inclusive setup. This behavior aligns with expectations based on decay-only amplitude arguments~\cite{Ballestrero:2017bxn}. With the application of fiducial cuts, interference effects remain below the percent level --- relative to the unpolarized cross section --- for both the SM and the squared dimension-six contributions, independent of their CP properties. In contrast, the CP-even linear term exhibits a more pronounced interference between longitudinal and transverse polarizations, reaching approximately $4\%$. Remarkably, the~interference between the SM amplitude and the CP-odd BSM contribution results in polarization interference effects comparable in size to the total unpolarized cross section. These significant shifts in the~interference pattern become even more evident when examining differential distributions of the azimuthal decay angles.

{\renewcommand{\arraystretch}{1.5}
\begin{table}[!t]
\begin{center}
\begin{tabular}{|c|ccc|ccc|}
\hline
& \multicolumn{3}{c|}{$\sigma$ [fb], inclusive, \cref{eq:inclusive}}& \multicolumn{3}{c|}{$\sigma$ [fb], ATLAS fiducial, \cref{eq:fiducial}}\\
\hline
& unpolarized & $W_L$ & $W_T$ & unpolarized & $W_L$ & $W_T$ \\ 
\hline
SM & \phantom{+}$97.27(1)$ & \phantom{+}$17.660 (3) $ & \phantom{+}$79.63(1)$ & \phantom{+}$34.64(3)$ & \phantom{+}$7.310(3)$ & \phantom{+}$27.15(2)$\\ \hline
$Q_W $ (lin.) & $-1.388 (1)$ & \phantom{+}$0.2758 (2)$ & $-1.6642 (8)$ & $-0.975 (1)$ & \phantom{+}$0.1313 (2)$ & $-1.152 (1)$\\ 
$Q_W $ (quad.) & \phantom{+}$12.076 (2)$ & \phantom{+}$0.1870 (0)$ & \phantom{+}$11.890 (2)$ & \phantom{+}$6.300 (2)$ & \phantom{+}$0.0803 (0)$ & \phantom{+}$6.203 (2)$\\ \hline
$Q_{\widetilde{W}}$ (lin.) & \phantom{+}$0.0071 (9)$ & $-0.0081 (1)$ & \phantom{+}$0.0160 (7)$ & $-0.013 (1)$ & $-0.0041 (1)$ & \phantom{+}$0.0157 (9)$\\ 
$Q_{\widetilde{W}}$ (quad.) & \phantom{+}$12.419 (2)$ & \phantom{+}$0.3013 (0)$ & \phantom{+}$12.118 (2)$ & \phantom{+}$6.444 (2)$ & \phantom{+}$0.1312 (1)$ & \phantom{+}$6.293 (2)$\\ 
\hline
$10^2 \cdot Q_{HW\!B}$ (lin.)& $ -4.180(1) $ & $ -3.977(1) $ & $ -0.2024(4) $ & $ -2.1102(7) $ & $ -2.1172(6) $ & $ 0.0124(3) $ \\ 
$10^2 \cdot Q_{HW\!B}$ (quad.) & $ 0.06683(1) $ & $ 0.06053(1) $ & $ 0.00630(0) $ & $ 0.03588(1) $ & $ 0.03306(1) $ & $ 0.00262(0) $ \\ 
\hline
$10^5 \cdot Q_{H\widetilde{W}\!B}$ (lin.) & $ 2.6(1) $ & $ 3.79(4) $ & $ -1.18(4) $ & $ 8.2(1) $ & $ 1.94(2) $ & $ -0.60(3) $ \\
$10^5 \cdot Q_{H\widetilde{W}\!B}$ (quad.) & 0.7119(1) & 0.6001(1) & 0.1118(1) & 0.3166(1) & 0.2676(1) & 0.04548(1) \\ 
\hline
\end{tabular}\qquad
\end{center}
\vspace{0mm}
\caption{Inclusive and fiducial NLO QCD cross sections in units of $\rm fb$ for $W^+ Z$ production at $\sqrt{s} = 13 \, {\rm TeV}$, considering (un)polarized $W$ bosons and unpolarized $Z$. All predictions are obtained in the DPA. The labelling and notation follow the conventions introduced in~\cref{tab:comp}. The quoted BSM cross sections are obtained using~$C_W/\Lambda^2 = 1 \, {\rm TeV}^{-2}$, $C_{\widetilde{W}}/\Lambda^2 = 1 \, {\rm TeV}^{-2}$, $C_{HW\!B}/\Lambda^2 = 1 \cdot 10^{-2} \, {\rm TeV}^{-2}$, and $C_{H\widetilde{W}\!B}/\Lambda^2 = 1 \cdot 10^{-2} \, {\rm TeV}^{-2}$. Note that, as indicated, all contributions from the~$Q_{HW\!B}$ operator are rescaled by a factor of $10^2$, while those from the $Q_{H\widetilde{W}\!B}$ operator are rescaled by a factor of $10^5$. \label{tab:Wpol}}
\end{table}
}

\begin{figure}[t!]
\centering
\includegraphics[width=.49\textwidth,page=1]{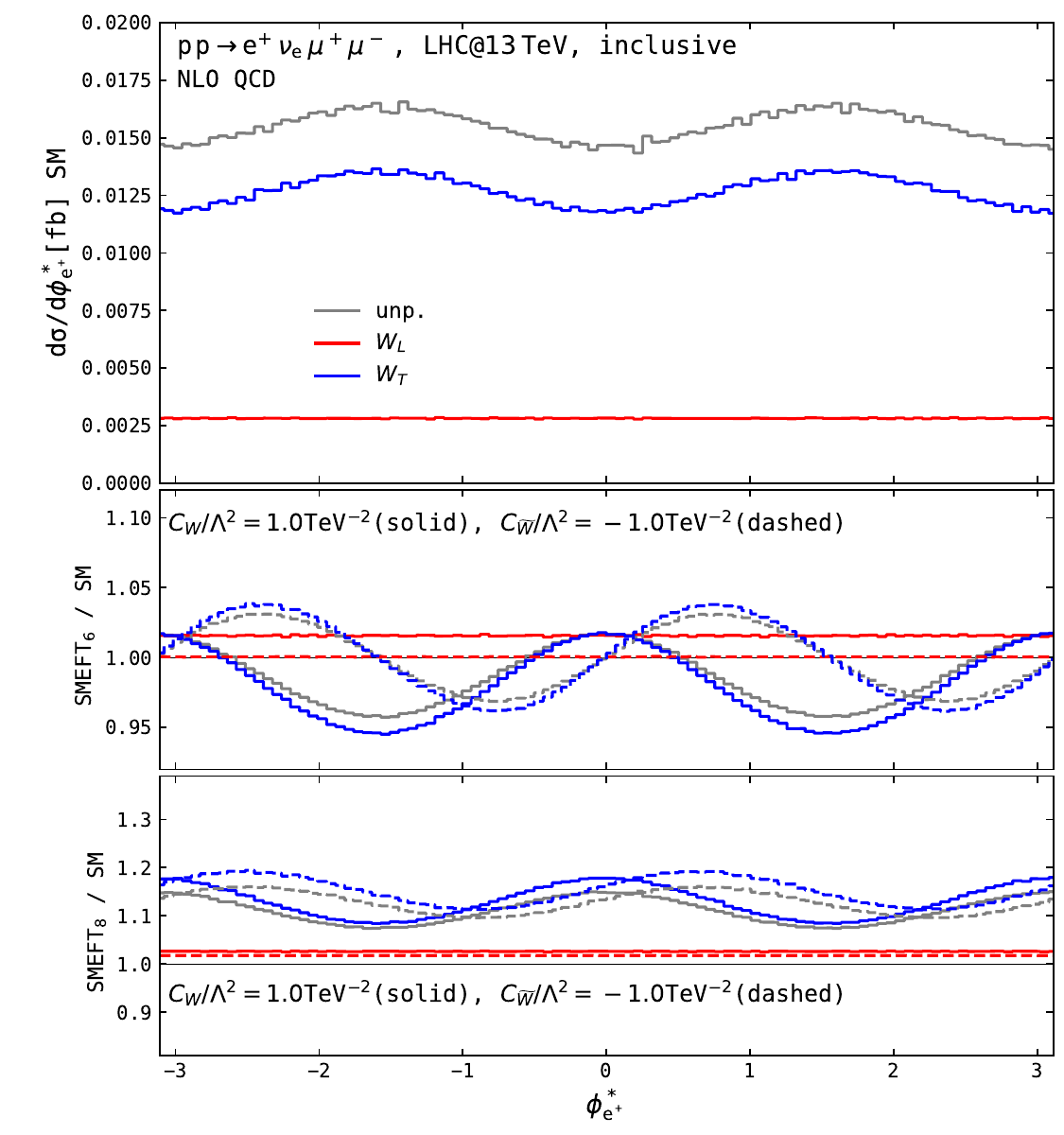}
\includegraphics[width=.49\textwidth,page=2]{w_pol.pdf}
\vspace{2mm}
\caption{Differential NLO QCD distributions at $\sqrt{s}= 13 \, {\rm TeV}$ are shown for the positron azimuthal decay angle in the $W$-boson rest frame ($\phi^\ast_{e^+}$), considering a longitudinal ($L$, red), transverse ($T$, blue), or unpolarized (unp., gray) $W$ boson, while the $Z$ boson remains unpolarized. The left and right panels correspond to the inclusive and fiducial setups, as defined in \cref{eq:inclusive,eq:fiducial}, respectively. Each panel is divided into three sections. The upper section displays the absolute SM predictions, while the lower two sections show the ratios of BSM to SM results. The BSM curves correspond to $C_W/\Lambda^2 = 1 \, {\rm TeV}^{-2}$ (solid lines) and $C_{\widetilde{W}}/\Lambda^2 = 1 \, {\rm TeV}^{-2}$ (dashed lines), with $\text{SMEFT}_6$ and $\text{SMEFT}_8$ defined according to~\cref{fig:comp}.}\label{fig:WZ_lW}
\end{figure}

\cref{fig:WZ_lW} displays the differential distributions of the positron decay angle in the $W$-boson rest frame~($\phi^\ast_{e^+}$) for both the inclusive and fiducial setups. This observable is particularly important due to its ability to further ``resurrect'' dimension-six interference effects and its pronounced sensitivity to CP-odd contributions~\cite{Azatov:2017kzw,Panico:2017frx,Azatov:2019xxn,Baglio:2019uty,ElFaham:2024uop}. In the inclusive setup, the distributions for longitudinal $W$~bosons are flat, reflecting the absence of azimuthal dependence due to the helicity structure of both the SM and the dimension-six SMEFT amplitudes. In~contrast, the transverse polarization states exhibit interference between left- and right-handed components, introducing a~$\sin \hspace{0.25mm}(2\phi^\ast_{e^+})$ modulation in the SM with an amplitude of around $7\%$ of the mean value. The~linear~$Q_W$ term reduces the amplitude of the modulation to about $3\%$ relative to the SM. The~corresponding quadratic term has little effect on the overall shape. The effects of~$Q_{\widetilde{W}}$ closely resemble those observed for~$Q_W$, with the main difference being that the modulation undergoes a phase shift of $\pi/4$. It is immediately apparent that fiducial cuts significantly distort the shape of the~$\phi^\ast_{e^+}$ distribution compared to the inclusive case, affecting both the SM baseline and the angular modulations induced by SMEFT operators. The most pronounced effects appear at $\phi^\ast_{e^+} = 0$, $\pm \pi$, which correspond to the minima of the angular spectrum. It is important to note that the inclusive and fiducial distributions shown in~\cref{fig:WZ_lW} are based on the reconstruction of the $W$-boson rest frame using the neutrino momentum from MC truth-level information. In realistic scenarios, the fiducial distributions would be further distorted by the challenges of single-neutrino reconstruction~\cite{ATLAS:2019bsc}. Nevertheless, studies have shown that even when accounting for both fiducial cuts and neutrino reconstruction, the sensitivity to CP-odd effects induced by the insertion of the $Q_{\widetilde{W}}$ operator remains largely intact~\cite{ElFaham:2024uop}.

Besides the least constrained operators $Q_W$ and $Q_{\widetilde{W}}$, we have also examined singly-polarized signals arising from insertions of the other operators listed in~\cref{eq:operators}, using the SMEFT benchmarks defined in \cref{eq:CPevenbench,eq:CPoddbench}. Our numerical analysis confirms that the Wilson coefficient $C_{H\widetilde{W}}$ does not contribute to $pp \to W^+ Z$ production, consistent with the discussion presented towards the end of \cref{sec:TGCs}. The last four rows of~\cref{tab:Wpol} display the linear and quadratic corrections to $W^+ Z$ production from the contributions of $Q_{HW\!B}$ and $Q_{H\widetilde{W}\!B}$ to~\cref{eq:LWWV}. Note that the operator~$Q_{HW\!B}$ also modifies the couplings $g_1$ and $g_2$ --- see~\cref{eq:LEPshifts,eq:GFshifts} --- and consequently affects the $W$- and $Z$-boson couplings to leptons and quarks, which enter both the production and decay stages of diboson processes. The associated scheme-dependent shifts are not incorporated in our new \POWHEGBOXRES{} implementation and are thus not included in the $Q_{HW\!B}$ results presented in~\cref{tab:Wpol}. As anticipated from their structure after~EWSB, the operators $Q_{HW\!B}$ and $Q_{H\widetilde{W}\!B}$ primarily affects the longitudinal polarization of the final-state $W$~boson, in~contrast to~$Q_W$ and~$Q_{\widetilde{W}}$, which predominantly impact its transverse polarization. The linear contributions from the CP-even operator $Q_{HW\!B}$ are suppressed by three orders of magnitude compared to the SM predictions. The~quadratic contributions are even more suppressed due to the smallness of the corresponding Wilson coefficient, rendering them negligible from an experimental perspective. Similarly, both the linear and quadratic contributions from the CP-odd operator $Q_{H\widetilde{W}\!B}$ are suppressed by approximately six orders of magnitude relative to the SM cross sections. This suggests that, given existing bounds on $Q_{HW\!B}$ and $Q_{H\widetilde{W}\!B}$ --- for example, those derived in~\cref{sec:Higgsphysics} from~$h \to \gamma \gamma$ --- the effects of these operators on diboson production at the LHC are too small to be experimentally detectable. We therefore do not consider their contributions further in this analysis.

We also examined the individual polarization states of the $Z$ boson produced alongside an unpolarized $W$ boson. However, we do not present these results here, as they do not reveal additional significant insights beyond those already demonstrated by signals involving a polarized $W$ boson. 

\subsection{Doubly-polarized signals} \label{sec:wzpol}

In this section, we address doubly-polarized signals, which constitute the primary modeling inputs for polarization analyses performed by ATLAS and CMS using LHC~Run~2 data~\cite{CMS:2020etf,ATLAS:2022oge,ATLAS:2023zrv,ATLAS:2024qbd}. As~previously discussed, the operator $Q_W$ modifies the triple-gauge-boson vertices $WWZ$ and $WW\gamma$. Within the DPA, only the left diagram in~\cref{fig:WZ_eft_diag} with an insertion of $Q_W$ contributes. By making the Lorentz structure of the SMEFT vertex explicit, the production-level contribution can be analyzed for specific $W$ and $Z$ boson polarization states $\lambda_W$ and $\lambda_Z$:
\beq \label{eq:AQW}
\begin{split}
{\cal A}_{\lambda_W \lambda_Z}(Q_W) & \propto \frac{C_W}{\Lambda^2}\, \Bigg \{\, g_{{\rho}{\mu}}\left [ - m_Z^2  \, p_{1 \hspace{0.25mm}\nu} + \frac{s - m_W^2 - m_Z^2}{2} \, p_{2 \hspace{0.25mm}\nu} \right ] \\[2mm]
& \hspace{1.5cm}+ g_{\rho \nu}\left[ -\frac{s- m_W^2- m_Z^2}{2} \, p_{1 \hspace{0.25mm}\mu} + m_W^2 \, p_{2 \hspace{0.25mm}\mu} \right ] \\[2mm]
& \hspace{1.5cm}+ g_{\mu \nu}\left[ \hspace{0.5mm} \frac{s - m_W^2 + m_Z^2}{2} \, p_{1 \hspace{0.25mm}\rho} - \frac{s + m_W^2 - m_Z^2}{2} \, p_{2 \hspace{0.25mm}\rho}  \right ] \\[2mm]
& \hspace{1.5cm}+ p_{1 \hspace{0.25mm}\nu}\hspace{0.5mm}p_{2 \hspace{0.25mm}\rho}\hspace{0.5mm}p_{12 \hspace{0.25mm}\mu}- p_{1 \hspace{0.25mm}\rho}\hspace{0.5mm}p_{2 \hspace{0.25mm}\mu}\hspace{0.5mm}p_{12 \hspace{0.25mm}\nu}\Bigg \}\; {\cal P}^{\rho}(p_{12}) \, \varepsilon_{\lambda_W}^{\mu}(p_1) \, \varepsilon_{\lambda_Z}^{\nu}(p_2) \,.
\end{split}
\eeq
Here, $p_1^\mu$ ($p_2^\nu$) represents the four-momentum of the $W$ ($Z$) boson, associated with the polarization vector $\varepsilon_{\lambda_W}^\mu (p_1)$ $\big($$\varepsilon_{\lambda_Z}^\nu (p_2)$$\big)$. The tensor ${\cal P}^{\rho}(p_{12})$ is the amputated amplitude for the production of an off-shell $W$ boson with four-momentum $p_{12}= p_1 + p_2$. 

{\renewcommand{\arraystretch}{1.5}
\begin{table}[!t]
\begin{center}
\begin{tabular}{|cccccc|}
\hline
& \multicolumn{4}{c}{$\sigma$ [fb], inclusive, \cref{eq:inclusive}}& \\
\hline
contribution & unpolarized & $W_L Z_L$ & $W_L Z_T$ & $W_T Z_L$ & $W_T Z_T$ \\
\hline
SM & \phantom{+}$97.28(2)$ & \phantom{+}$4.501(1)$ & \phantom{+}$13.161(3) $ & \phantom{+}$12.733(3)$ & \phantom{+}$66.89(1)$ \\ 
\hline
$Q_W $ (lin.) & $-1.388(1)$ & 0 & \phantom{+}$0.2760(2)$ & \phantom{+}$0.3344(3)$ & $-1.9982(7)$ \\
$Q_W $ (quad.) & \phantom{+}$12.077(2)$ & 0 & \phantom{+}$0.1870(0)$ & \phantom{+}$0.2405(0)$ & \phantom{+}$11.649(2)$ \\ 
\hline
$Q_{\widetilde{W}}$ (lin.) & \phantom{+}$0.007(1)$ & 0 & $-0.0082(1)$ & \phantom{+}$0.0121(1)$ & \phantom{+}$0.0032(7)$ \\
$Q_{\widetilde{W}}$ (quad.) & \phantom{+}$12.419(2)$ & 0 & \phantom{+}$0.3013(0)$ & \phantom{+}$0.3548(1)$ & \phantom{+}$11.763(2)$ \\
\hline \hline
& \multicolumn{4}{c}{$\sigma$ [fb], ATLAS fiducial, \cref{eq:fiducial}}& \\
\hline
contribution & unpolarized & $W_L Z_L$ & $W_L Z_T$ & $W_T Z_L$ & $W_T Z_T$ \\
\hline
SM & \phantom{+}$34.64(2)$ & \phantom{+}$1.968(1)$ & \phantom{+}$5.357(2)$ & \phantom{+}$5.100(2)$ & \phantom{+}$21.99(2)$ \\
\hline
$Q_W$ (lin.) & $-0.975(1)$ & 0 & \phantom{+}$0.1229(3)$ & \phantom{+}$0.1435(3)$ & $-1.264(1)$ \\
$Q_W $ (quad.) & \phantom{+}$6.299(2)$ & 0 & \phantom{+}$0.0803(0)$ & \phantom{+}$0.1001(0)$ & \phantom{+}$6.110(2)$ \\ 
\hline
$Q_{\widetilde{W}}$ (lin.) & $-0.013(1)$ & 0 & $-0.0039(1)$ & \phantom{+}$0.0056(1)$ & \phantom{+}$0.0102(9)$ \\
$Q_{\widetilde{W}}$ (quad.)& \phantom{+}$6.443(2)$ & 0 & \phantom{+}$0.1312(1)$ & \phantom{+}$0.1661(1)$ & \phantom{+}$6.132(2)$ \\
\hline
\end{tabular}\qquad
\end{center}
\vspace{0mm}
\caption{Inclusive and fiducial NLO QCD cross sections in units of $\rm fb$ for $W^+ Z$ production at $\sqrt{s}= 13 \, {\rm TeV}$ with doubly-polarized final states, computed within the DPA. The notation follows that of~\cref{tab:Wpol}. The~reported BSM cross sections are obtained using $C_W/\Lambda^2 = 1 \, {\rm TeV}^{-2}$ and $C_{\widetilde{W}}/\Lambda^2 = 1 \, {\rm TeV}^{-2}$.
\label{tab:doublypol}}
\end{table}

\begin{figure}[t!]
\centering
\includegraphics[width=.49\textwidth,page=1]{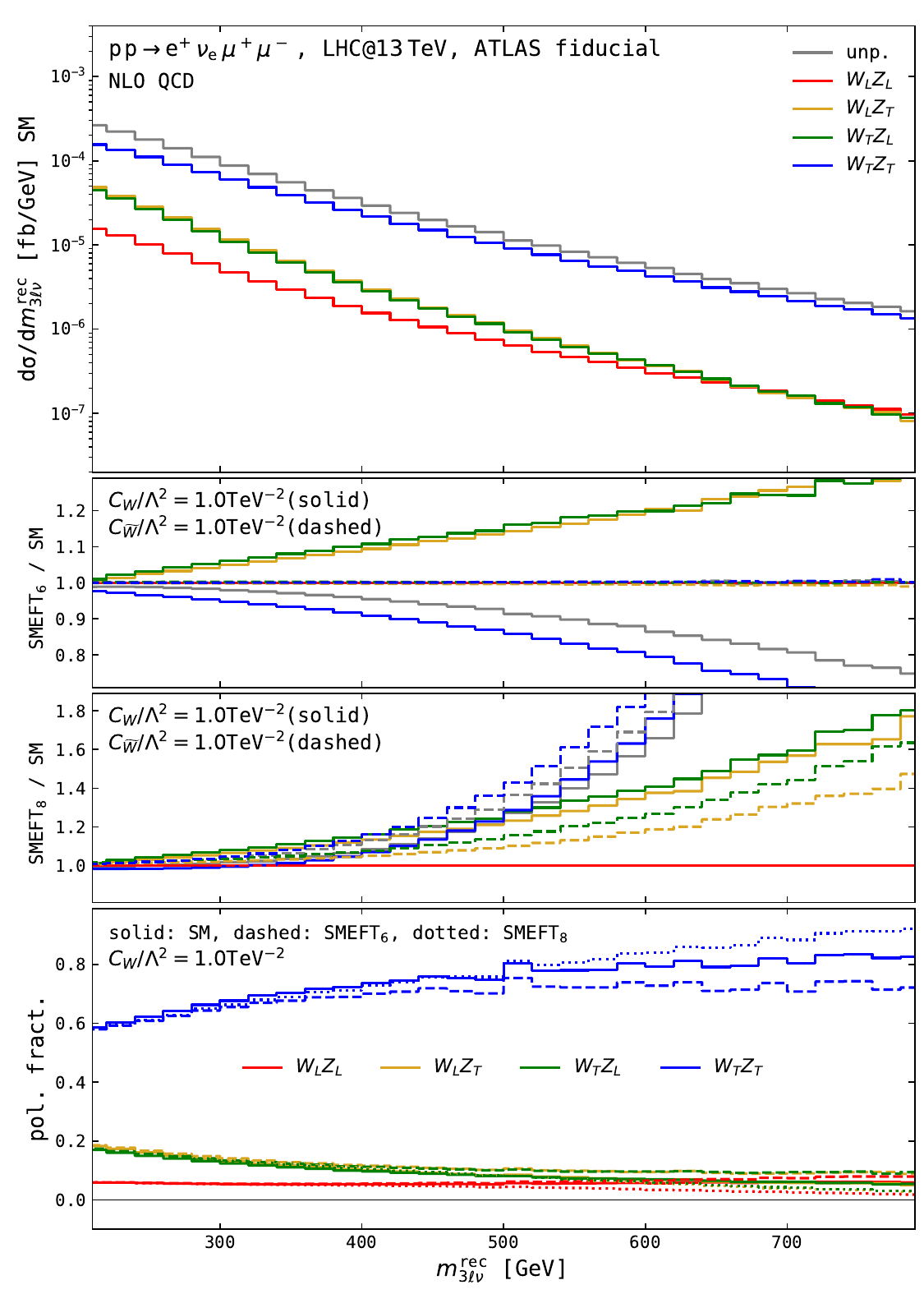}
\includegraphics[width=.49\textwidth,page=2]{doubly_polarized.pdf}
\vspace{2mm}
\caption{Differential NLO QCD distributions at $\sqrt{s} = 13 \, {\rm TeV}$ are presented for the reconstructed invariant mass of the diboson system~($m^{\rm rec}_{3\ell\nu}$) in the left panel, and in the rapidity separation between the positron and the $Z$ boson $\big($$|\Delta y_{e^+Z}|$$\big)$ in the right panel. The results are categorized by joint polarization configurations: $LL$~(red), $LT$~(yellow), $TL$~(green), and $TT$~(blue), where the first (second) label denotes the polarization of the $W$ ($Z$) boson. Unpolarized distributions~(gray) are included for comparison. The analysis is performed within the fiducial setup specified in~\cref{eq:fiducial}. The panel layout matches that of \cref{fig:WZ_lW} for the top three sections. In~the bottom panels, joint polarization fractions, defined as polarized cross sections normalized to the unpolarized ones, are shown. The BSM curves correspond to the choices $C_W/\Lambda^2 = 1 \, {\rm TeV}^{-2}$~(solid lines) and $C_{\widetilde{W}}/\Lambda^2 = 1 \, {\rm TeV}^{-2}$~(dashed lines) of Wilson coefficients, with $\text{SMEFT}_6$ and $\text{SMEFT}_8$ defined according to~\cref{fig:comp}.} \label{fig:diff_doublypol}
\end{figure}

In the diboson CM frame, where $\sqrt{s}$ is the total energy of the system, the longitudinal polarization vectors are expressed as linear combinations of the two boson momenta, with coefficients that depend on $m_W$, $m_Z$, and $s$. For simplicity, we only present their expansions in the high-energy limit $s \to \infty$. We find:
\beq \label{eq:polvec}
\begin{split}
\varepsilon^\mu_{L}(p_1) & = \frac{1}{m_W}\, p_1^\mu - \frac{2 \hspace{0.125mm}m_W}{s}\, p_2^\mu + \frac{2\hspace{0.125mm}m_W \hspace{0.25mm}m_Z^2}{s^2}\, p_1^\mu - \frac{2 \hspace{0.125mm}m_W \left (m_W^2 + m_Z^2 \right )}{s^2}\, p_2^\mu + {\cal O} \left(\frac1{s^3}\right) \,, \\[2mm]
\varepsilon^\nu_{L}(p_2) & = \frac{1}{m_Z}\, p_2^\nu - \frac{2 \hspace{0.125mm}m_Z}{s}\, p_1^\nu + \frac{2 \hspace{0.125mm}m_Z \hspace{0.25mm}m_W^2}{s^2}\, p_2^\nu - \frac{2 \hspace{0.125mm}m_Z \left (m_W^2+m_Z^2 \right )}{s^2}\, p_1^\nu + {\cal O} \left(\frac{1}{s^3}\right)\,.
\end{split}
\eeq
From~\cref{eq:AQW,eq:polvec}, one easily derives that the double-longitudinal SMEFT amplitude with a single insertion of the CP-even operator $Q_W$ vanishes at all energy scales, both at tree level and with QCD corrections included:
\beq \label{eq:ALLQW}
{\cal A}_{LL}(Q_W) = 0 \,.
\eeq
The same conclusion applies to an insertion of the CP-odd operator $Q_{\widetilde{W}}$. 

The above considerations suggest that the purely longitudinal signal receives contributions solely from the SM, which is confirmed by direct simulations of the doubly-polarized signals using our new combined \recolatwo{} and \POWHEGBOXRES{} code. \cref{tab:doublypol} presents the integrated cross sections for the doubly-polarized signals under both inclusive and fiducial conditions \cref{eq:inclusive,eq:fiducial}. In line with general expectations regarding the energy dependence of $2 \to 2$ scattering amplitudes \cite{Azatov:2017kzw,Azatov:2016sqh}, interference effects between the SM and the $Q_W$ operator --- amounting to approximately $3\%$ relative to the SM --- emerge for both purely transverse polarization states and mixed polarization configurations ($LT$ and $TL$), albeit with opposite signs. Relative to the linear term, the quadratic term yields a substantial enhancement of the $TT$ signal --- approximately $15\%$ --- due to the opening of the SM-suppressed $W_\pm Z_\pm$ helicity contribution. In~contrast, the mixed polarization channels receive quadratic contributions that are nearly comparable in magnitude to the linear ones. The linear CP-odd contributions from $Q_{\widetilde{W}}$ are at the sub-percent level, while the corresponding quadratic terms are comparable to the CP-even contributions for the $TT$ state and somewhat larger for the mixed polarization channels.

In \cref{fig:diff_doublypol}, we present differential distributions for two observables within the fiducial ATLAS setup defined in~\cref{eq:fiducial}. The reconstructed invariant mass of the diboson system $m_{3\ell\nu}^{\rm rec}$, shown in the left panel, illustrates how the energy dependence of different polarization states evolves in the~SM and in the presence of BSM effects. In particular, the contribution from the operator $Q_W$ to the~$TT$~amplitude exhibits a quadratic energy growth~\cite{Azatov:2017kzw}, which dominates the high-mass region of the unpolarized BSM signal --- both at the linear level and when including the squared SMEFT term. A similar energy scaling appears at quadratic order for the BSM prediction involving the CP-odd operator $Q_{\widetilde{W}}$, while its linear contribution remains suppressed across the entire mass range. Mixed polarization states, which are suppressed at high energies in the SM~\cite{Vayonakis:1976vz,Willenbrock:1987xz}, receive a linear-level enhancement from the CP-even $Q_W$ operator, owing to the linear growth of the dimension-six amplitude with the energy of the diboson system relative to the SM.

In the right panel of \cref{fig:diff_doublypol}, we examine the rapidity separation $|\Delta y_{e^+Z}|$ between the positron originating from the $W$-boson decay and the $Z$ boson, reconstructed from the antimuon-muon pair. This observable has demonstrated excellent sensitivity to different polarization states in the recent ATLAS analyses~\cite{ATLAS:2022oge,ATLAS:2024qbd}, owing to its strong correlation with the scattering angle between the two gauge bosons. At linear order, the mixed polarization states receive a BSM correction from $Q_W$ that is positive for $|\Delta y_{e^+Z}| \lesssim 1.7$ --- the most populated region --- and thus counteracts the negative linear correction affecting the $TT$ state. In the suppressed region with large rapidity separations, the linear contribution from $Q_W$ to the mixed states becomes negative. When including quadratic terms, the $TT$ contribution receives a substantial correction in the central region, while BSM effects are significantly reduced in the tails of the distribution. There, the negative linear correction to the~$LT$ and $TL$ states remains the dominant deviation from the SM.

The polarization fractions --- defined as doubly-polarized cross sections normalized to the unpolarized ones --- are also influenced by the SMEFT insertions. These effects are most pronounced not only in the kinematically suppressed tails of energy-dependent distributions, but also in the central regions of angular and rapidity distributions. For example, including both linear and quadratic $Q_W$~terms leads to a roughly $10\%$ increase in the $TT$ fraction at $|\Delta y_{e^+Z}| \simeq 0$ and for $m_{3\ell\nu}^{\rm rec} \gtrsim 700 \, {\rm GeV}$. Although the energy suppression in the transverse-momentum and invariant-mass distributions hampers the sensitivity to individual polarization states, the notable modifications in the angular distributions due to the SMEFT effects suggest that using template fits could enhance the precision of Wilson coefficient constraints compared to standard LHC analyses. Moreover, SMEFT-based modeling of polarized templates opens the door to a model-independent determination of polarization fractions, an approach previously restricted to the SM.

\begin{figure}[t!]
\centering
\includegraphics[width=.49\textwidth,page=2]{doubly_polarized_ps_ratio_nlops_nlo.pdf}
\includegraphics[width=.49\textwidth,page=1]{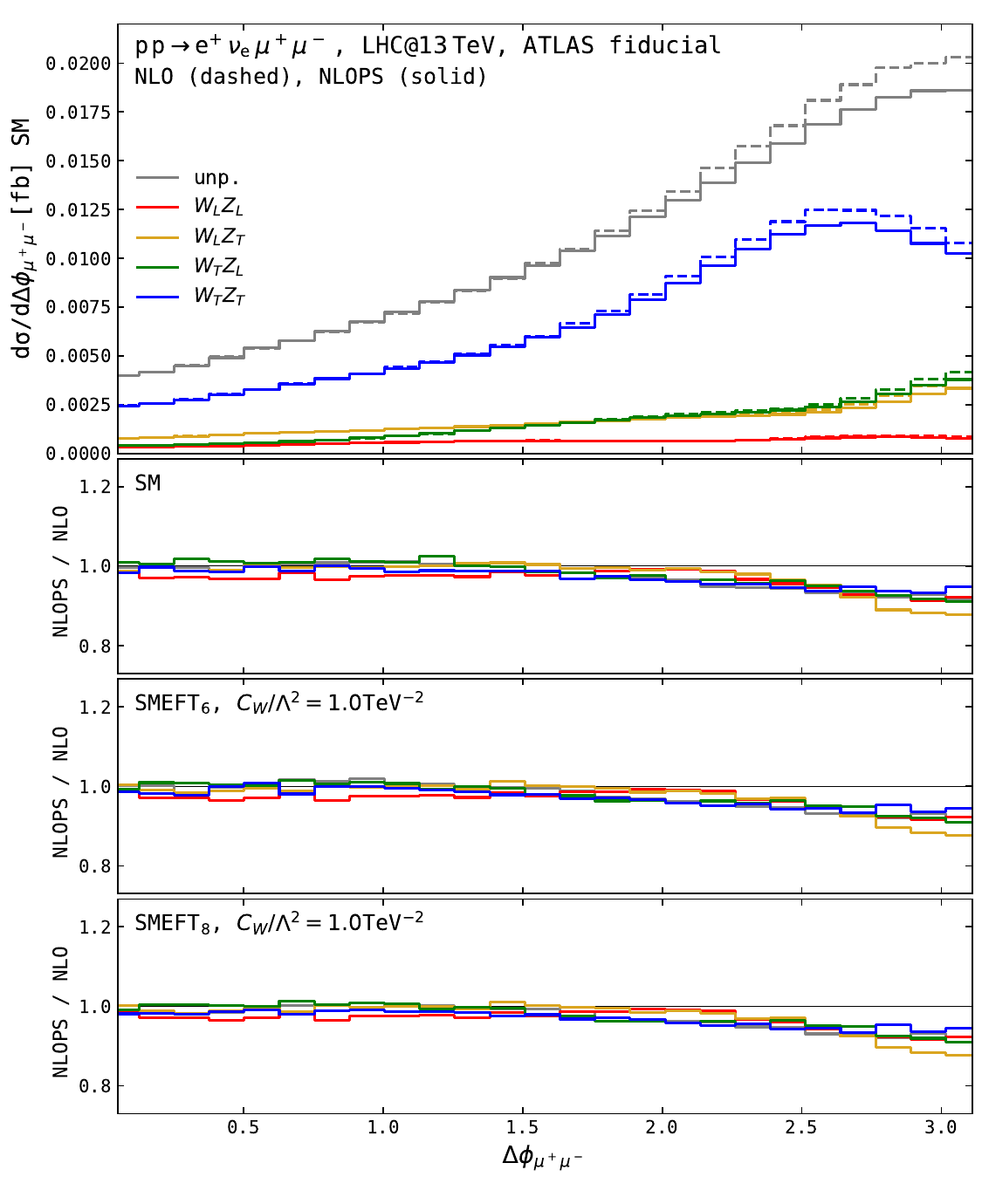}
\vspace{2mm}
\caption{Differential distributions at $\sqrt{s} = 13 \, {\rm TeV}$ are shown for the transverse momentum of the diboson system ($p_{T, WZ}$) in the left panel, and the azimuthal angle between the antimuon and muon~($\Delta \phi_{\mu^+ \mu^-}$) in the right panel. The observables are evaluated within the fiducial phase space defined in \cref{eq:fiducial}. Each panel is divided into four sections. The top section displays the absolute SM predictions at fixed-order NLO QCD (dashed lines) and with NLO+PS corrections (solid lines). The bottom three sections show the ratios of the NLO+PS to NLO results for both the SM and BSM predictions. The color scheme follows that of \cref{fig:diff_doublypol}. The displayed BSM predictions correspond to the Wilson coefficient choice $C_W/\Lambda^2 = 1 \, {\rm TeV}^{-2}$, with $\text{SMEFT}_6$ and $\text{SMEFT}_8$ defined as in \cref{fig:comp}.} \label{fig:nlops_diff}
\end{figure}

\subsection{Parton-shower effects} \label{sec:pseffects}

The presented \POWHEGBOXRES{} implementation enables the matching of NLO QCD predictions for polarized diboson processes with PS and hadronization effects, via a user-friendly interface to \noun{Pythia~8.2}, closely following the original SM setup described in~\citere{Pelliccioli:2023zpd}. While the SMEFT contributions generally have a minimal impact on the low transverse momentum~(Sudakov) region and instead predominantly affect high-energy kinematic regions described by the fixed-order calculation, their inclusion is exact. That is, SMEFT effects are consistently incorporated into the computation --- without any reweighting --- through their contribution to both the Sudakov form factor $\Delta (t)$ in~\cref{eq:sudakov} and the NLO-accurate event weights $\bar{B}(\tilde{\Phi}_4)$ in~\cref{eq:Bbar}.

In~\cref{fig:nlops_diff}, we compare fixed-order and PS-matched predictions for two differential observables, the transverse momentum of the diboson system ($p_{T,WZ}$) and the azimuthal angle between the antimuon and muon ($\Delta \phi_{\mu^+ \mu^-}$), incorporating BSM effects from the CP-even operator $Q_W$. As~shown in the figure, the overall impact of PS effects on the $\text{SMEFT}_6$ predictions involving the $Q_W$ operator closely resembles that observed in the SM. The inclusion of quadratic SMEFT contributions modifies this behavior for certain polarization configurations. This is illustrated in the left panel of~\cref{fig:nlops_diff}, where PS~effects suppress the $TT$ signal in the most populated region of the $p_{T,WZ}$ distribution. As a result, PS~effects similarly reduce the $\text{SMEFT}_8$ predictions for the unpolarized $p_{T,WZ}$ distribution relative to the~SM. The~SMEFT amplitude involving an insertion of the operator $Q_W$ allows only left-left and right-right helicity states (with polarizations defined in the diboson rest frame). Therefore, as in the case of the Higgs decays to two EW gauge bosons, the diboson system predominantly forms a spin-singlet state, which is typically produced with negligible transverse momentum~\cite{Pelliccioli:2023zpd}. Real~QCD~radiation, both at fixed order and after PS matching, introduces helicity configurations absent at LO, mainly due to the opening of the gluon-quark partonic channel. This provides a qualitative understanding of the low-$p_{T, WZ}$ behavior of the dominant $TT$ signal. A similar pattern was observed in the unpolarized process and confirmed through comparison with the off-shell \SMEFTNLO{} and \MadGraph{} results presented in~\cite{ElFaham:2024uop}. However, fixed-order NLO predictions are clearly unreliable in the low-$p_{T, WZ}$ region, highlighting the need for higher-order QCD corrections, either via PS matching or through an all-order resummed calculation. In contrast, for the azimuthal angle difference $\Delta \phi_{\mu^+ \mu^-}$ displayed in the right panel, PS effects influence both the SMEFT and SM predictions similarly, showing minimal variation across the four polarization configurations and exhibiting negligible sensitivity to the SMEFT expansion order.

While not explicitly shown, we emphasize that PS-matched predictions have been generated with the new MC code for all relevant operators contributing to the triple-gauge-boson interactions defined in~\cref{eq:LWWV}. From~a technical standpoint, we found that event generation with SMEFT contributions is numerically more stable if the linear dimension-six terms are included alongside the~SM~contributions, rather than handled separately. This arises from the fact that the linear~SMEFT terms are not positive definite, which can otherwise cause instabilities at the level of unweighted-event generation. 

\subsection{Quantum tomography} \label{sec:tomography}

The new calculation presented in this article enables a more precise determination of spin-correlation coefficients associated with EW gauge bosons. These coefficients serve as essential inputs for the analysis of spin entanglement and Bell non-locality observables in inclusive diboson production processes~\cite{Grossi:2024jae,Aoude:2023hxv,Barr:2022wyq,Fabbrichesi:2023cev,Grabarczyk:2024wnk,Afik:2025ejh,Goncalves:2025qem}. Within the DPA approach, unpolarized $W^+ Z$ production with opposite-flavor leptonic decays $W^+ \to e^+ \nu_e$ and $Z \to \mu^+ \mu^-$ can be unambiguously described using standard tree-level quantum tomography~\cite{Rahaman:2021fcz}, in the inclusive setup defined by \cref{eq:inclusive} and in the presence of QCD radiative corrections. Adopting the notation of~\citere{Grossi:2024jae}, the complete angular dependence of the diboson decay is expressed as:
\beq \label{eq:VV}
\begin{split}
\frac{d^5 \sigma}{d \Omega^\ast_{e^+}\hspace{0.125mm}d \Omega^\ast_{\mu^+}\hspace{0.125mm}d {\cal O}} & = \Bigg [ \frac{1}{(4\pi)^2}+ \frac{1}{4\pi}\, \sum_{l=1}^2 \sum_{m=-l}^{l}\, \alpha^{(W)}_{lm}({\cal O}) \, Y_{lm}\big (\theta^\ast_{e^+}, \phi^\ast_{e^+}\big ) \\[2mm]
& \phantom{xxx}+ \frac{1}{4\pi}\, \sum_{l=1}^2 \sum_{m=-l}^{l}\, \alpha^{(Z)}_{lm}({\cal O}) \, Y_{lm}\big (\theta^\ast_{\mu^+},\phi^\ast_{\mu^+}\big )
\\[2mm]
& \phantom{xxx}+ \sum_{l=1}^2 \sum_{l^\prime=1}^2 \sum_{m=-l}^{l}\sum_{m^\prime=-l^\prime}^{l^\prime}\, \gamma_{lm l^\prime m^\prime}({\cal O}) \, Y_{lm}\big (\theta^\ast_{e^+}, \phi^\ast_{e^+}\big ) \, Y_{l^\prime m^\prime}\big (\theta^\ast_{\mu^+}, \phi^\ast_{\mu^+}\big ) \Bigg] \, \frac{d \sigma}{d {\cal O}}\,.
\end{split}
\eeq 
Here, $d \Omega^\ast_{\ell^+}= d\cos\theta^\ast_{\ell^+} \hspace{0.5mm} d\phi^\ast_{\ell^+}$, where $\theta^\ast_{\ell^+}$ and $\phi^\ast_{\ell^+}$ denote the polar and azimuthal decay angles of the positively charged leptons, respectively, defined in the rest frame of the corresponding boson. The~spin quantization axis of the EW gauge bosons is taken along the trajectory of the diboson system in the laboratory frame. In \cref{eq:VV}, $Y_{lm}(\theta, \phi)$ denote the real-valued spherical harmonics, and~${\cal O}$~represents a generic observable that is independent of the four decay angles --- for example, the transverse momentum of the $Z$ boson. 

{\renewcommand{\arraystretch}{1.5}
\begin{table}[!t]
\begin{center}
\begin{tabular}{|c|r|rr|rr|}
\hline
\multicolumn{5}{|c}{\hspace*{3cm}fixed-order NLO QCD}& \\
\hline
coefficient & SM & $Q_W$ ($\text{SMEFT}_6$) & $Q_W$ ($\text{SMEFT}_8$) & $Q_{\widetilde{W}}$ ($\text{SMEFT}_6$) & $Q_{\widetilde{W}}$ ($\text{SMEFT}_8$) \\
\hline
$\alpha^{(W)}_{10}$ & $-0.0473^{+3}_{-1}$ & $-0.0472^{+3}_{-1}$ & $-0.0419^{+8}_{-5}$ & $-0.0474^{+3}_{-2}$ & $-0.0420^{+8}_{-5}$ \\ 
$\alpha^{(W)}_{20}$ & $0.0287^{+6}_{-2}$ & $0.0277^{+6}_{-2}$ & $0.0313^{+13}_{-7}$ & $0.0287^{+7}_{-2}$ & $0.0321^{+13}_{-7}$ \\ 
$\alpha^{(W)}_{22}$ & $0$ & $0$ & $0$ & $0.0067^{+4}_{-1}$ & $0.0059^{+2}_{-1}$ \\ 
$\alpha^{(W)}_{2-2}$ & $-0.0120^{+3}_{-4}$ & $-0.0053^{+7}_{-6}$ & $-0.0046^{+7}_{-6}$ & $-0.0121^{+3}_{-4}$ & $-0.0109^{+5}_{-5}$ \\ 
$\alpha^{(Z)}_{10}$ & $0.0040^{+4}_{-1}$ & $0.0042^{+4}_{-1}$ & $0.0038^{+3}_{-2}$ & $0.0040^{+4}_{-2}$ & $0.0035^{+4}_{-1}$ \\ 
$\alpha^{(Z)}_{20}$ & $0.0295^{+6}_{-2}$ & $0.0283^{+6}_{-1}$ & $0.0318^{+12}_{-6}$ & $0.0295^{+6}_{-2}$ & $0.0327^{+12}_{-6}$ \\ 
$\alpha^{(Z)}_{22}$ & $0$ & $0$ & $0$ & $0.0069^{+4}_{-1}$ & $0.0061^{+2}_{-1}$ \\ 
$\alpha^{(Z)}_{2-2}$ & $-0.0079^{+2}_{-3}$ & $-0.0011^{+6}_{-3}$ & $-0.0008^{+5}_{-4}$ & $-0.0077^{+2}_{-2}$ & $-0.0069^{+3}_{-4}$ \\ 
$\gamma_{1010}$ & $-0.0053^{+1}_{-2}$ & $-0.0056^{+0}_{-2}$ & $-0.0036^{+3}_{-4}$ & $-0.0053^{+1}_{-2}$ & $-0.0033^{+3}_{-4}$ \\ 
$\gamma_{2020}$ & $0.0013^{+1}_{-0}$ & $0.0012^{+1}_{-0}$ & $0.0015^{+1}_{-0}$ & $0.0013^{+1}_{-0}$ & $0.0016^{+0}_{-1}$ \\ 
\hline \hline
 \multicolumn{5}{|c}{\hspace*{3cm}NLO QCD + PS (QCD+QED) + hadronization + MPI}& \\
\hline
coefficient & SM & $Q_W$ ($\text{SMEFT}_6$) & $Q_W$ ($\text{SMEFT}_8$) & $Q_{\widetilde{W}}$ ($\text{SMEFT}_6$) & $Q_{\widetilde{W}}$ ($\text{SMEFT}_8$) \\
\hline
$\alpha^{(W)}_{10}$ & $-0.0483^{+3}_{-1}$ & $-0.0482^{+2}_{-1}$ & $-0.0426^{+9}_{-6}$ & $-0.0484^{+3}_{-1}$ & $-0.0427^{+9}_{-6}$ \\ 
$\alpha^{(W)}_{20}$ & $0.0298^{+6}_{-2}$ & $0.0288^{+6}_{-2}$ & $0.0324^{+13}_{-6}$ & $0.0298^{+6}_{-2}$ & $0.0333^{+12}_{-7}$ \\ 
$\alpha^{(W)}_{22}$ & $0$ & $0 $ & $0 $ & $0.0067^{+4}_{-1}$ & $0.0059^{+2}_{-1}$ \\ 
$\alpha^{(W)}_{2-2}$ & $-0.0115^{+3}_{-4}$ & $-0.0049^{+7}_{-5}$ & $-0.0042^{+7}_{-5}$ & $-0.0116^{+3}_{-4}$ & $-0.0104^{+5}_{-6}$ \\ 
$\alpha^{(Z)}_{10}$ & $0.0038^{+4}_{-1}$ & $0.0041^{+4}_{-1}$ & $0.0036^{+4}_{-1}$ & $0.0038^{+4}_{-1}$ & $0.0034^{+4}_{-2}$ \\ 
$\alpha^{(Z)}_{20}$ & $0.0292^{+5}_{-2}$ & $0.0280^{+6}_{-1}$ & $0.0315^{+12}_{-6}$ & $0.0291^{+6}_{-1}$ & $0.0323^{+12}_{-5}$ \\ 
$\alpha^{(Z)}_{22}$ & $0 $ & $0$ & $0$ & $0.0069^{+4}_{-1}$ & $0.0061^{+2}_{-1}$ \\ 
$\alpha^{(Z)}_{2-2}$ & $-0.0081^{+2}_{-2}$ & $-0.0012^{+6}_{-4}$ & $-0.0010^{+6}_{-3}$ & $-0.0079^{+2}_{-2}$ & $-0.0071^{+3}_{-3}$ \\ 
$\gamma_{1010}$ & $-0.0052^{+0}_{-2}$ & $-0.0056^{+0}_{-2}$ & $-0.0036^{+4}_{-4}$ & $-0.0052^{+0}_{-2}$ & $-0.0033^{+4}_{-4}$ \\ 
$\gamma_{2020}$ & $0.0013^{+1}_{-0}$ & $0.0012^{+1}_{-0}$ & $0.0015^{+1}_{-0}$ & $0.0013^{+1}_{-0}$ & $0.0016^{+1}_{-0}$ \\ 
\hline
\end{tabular}\qquad
\end{center}
\vspace{0mm}
\caption{Inclusive NLO QCD predictions for a set of relevant spin-correlation coefficients in $W^+ Z$ production at $\sqrt{s} = 13 \, {\rm TeV}$, as defined in \cref{eq:VV}. The upper and lower blocks present our results at fixed order and matched to full PS effects, respectively. The BSM predictions are obtained using $C_W/\Lambda^2 = 1 \, {\rm TeV}^{-2}$ and $C_{\widetilde{W}}/\Lambda^2 = 1 \, {\rm TeV}^{-2}$, with $\text{SMEFT}_6$ and $\text{SMEFT}_8$ defined as in~\cref{fig:comp}. \label{tab:qtom}}
\end{table}
}
Higher-order QCD corrections are known to significantly alter the LO values of the angular coefficients, due to the emergence of new helicity configurations and polarization-interference effects starting at NLO QCD~\cite{Grossi:2024jae}. Moreover, as recently shown~\cite{Grossi:2024jae,Goncalves:2025qem,DelGratta:2025qyp}, the decomposition in~\cref{eq:VV} becomes partially ambiguous when NLO EW corrections are included, which complicates the interpretation of these coefficients in terms of spin correlations. This ambiguity is primarily caused by loop-induced effects in the decays of the EW gauge bosons, while photon radiation from final-state particles plays a comparatively minor role. SMEFT effects in diboson processes have been previously studied within the framework of quantum tomography, though such analyses have so far been limited to~LO~accuracy~\cite{Aoude:2023hxv}. The results presented below represent a significant advancement in the precision of spin-correlation extraction, achieving for the first time NLO QCD accuracy matched to PS effects for both the SM and the dimension-six SMEFT operators affecting triple-gauge-boson couplings. The~coefficients $\alpha^{(W)}_{lm}$, $\alpha^{(Z)}_{lm}$, and $\gamma_{lm l^\prime m^\prime}$ corresponding to the fully inclusive cross section are extracted by projecting \cref{eq:VV} onto spherical harmonics: 
\beq \label{eq:spherharm}
\begin{split}
\alpha^{(W)}_{lm} & = \frac{1}{\sigma}\int \! d \Omega^\ast_{e^+}\int \! d \Omega^\ast_{\mu^+}\, \frac{d^4 \sigma}{d \Omega^\ast_{e^+}\,d \Omega^\ast_{\mu^+}}\, Y_{lm}\big (\theta^\ast_{e^+}, \phi^\ast_{e^+}\big ) \,, \\[2mm]
\alpha^{(Z)}_{lm}& = \frac{1}{\sigma}\int \! d \Omega^\ast_{e^+}\int \! d \Omega^\ast_{\mu^+}\, \frac{d^4 \sigma}{d \Omega^\ast_{e^+}\,d \Omega^\ast_{\mu^+}}\, Y_{l m}\big (\theta^\ast_{\mu^+}, \phi^\ast_{\mu^+}\big ) \,, \\[2mm]
\gamma_{lm l^\prime m^\prime} & = \frac{1}{\sigma}\int \! d \Omega^\ast_{e^+}\int \! d \Omega^\ast_{\mu^+}\, \frac{d^4 \sigma}{d \Omega^\ast_{e^+}\,d \Omega^\ast_{\mu^+}}\, Y_{lm}\big (\theta^\ast_{e^+}, \phi^\ast_{e^+}\big ) \, Y_{l^\prime m^\prime}\big (\theta^\ast_{\mu^+}, \phi^\ast_{\mu^+}\big ) \,. 
\end{split}
\eeq
It is important to note that in the above expressions, the dependence on the Wilson coefficients appears in both the numerators and denominators, potentially leading to more complex effects than those typically observed in standard cross sections and kinematic distributions. Moreover, the extraction of the coefficients from~\cref{eq:VV}, as defined in~\cref{eq:spherharm}, is only well-defined in a fully inclusive setup, i.e., without applying selection cuts on individual decay products~\cite{Grossi:2024jae}. Consequently, their determination is affected by systematic uncertainties stemming from the extrapolation from the fiducial to the inclusive phase-space region. For this reason, we present results within the inclusive setup specified in~\cref{eq:inclusive}.

A full analysis of all 80 independent coefficients in~\cref{eq:VV} lies beyond the scope of this work. Instead, we focus on a subset of coefficients that are phenomenologically relevant for inclusive diboson processes. This selection corresponds to those presented in~\citere{Grossi:2024jae}, with the addition of single-boson coefficients sensitive to CP-odd effects. \cref{tab:qtom} presents our NLO~QCD results for the inclusive spin-correlation coefficients of interest, shown at fixed order (upper block) and after matching to the full PS effects using \noun{Pythia 8.2}, including QCD and QED showering, hadronization, and~MPI (lower block). The quoted uncertainties stem from the standard seven-point correlated variations of the QCD renormalization and factorization scales, and are at the percent level. It is worth noting that in the SM, the coefficients $\alpha^{(V)}_{22}$ for $V = W, Z$ vanish identically. The linear contribution from the $Q_W$ operator significantly alters $\alpha^{(V)}_{2-2}$, while its quadratic contribution induces substantial deviations across all SM coefficients. In contrast, the linear and quadratic contribution from the $Q_{\widetilde{W}}$ operator generates a non-zero~$\alpha^{(V)}_{22}$, consistent with the $\sin \hspace{0.25mm}(2\phi^\ast_{e^+})$ modulation observed on the left-hand side in~\cref{fig:WZ_lW}. For the $\alpha^{(V)}_{l0}$ coefficients, the squared SMEFT contributions approximately double the QCD scale uncertainties observed in the~SM. In~the PS-matched predictions, the dominant effect arises from the QED shower, which resums higher-order photon emissions from both initial-state and final-state particles. However, the~$\alpha^{(V)}_{lm}$ and $\gamma_{lm l^\prime m^\prime}$ coefficients show deviations from their fixed-order counterparts that are largely encompassed by the QCD scale uncertainties. A~complete characterization of the spin-density matrix for the $W^+ Z$ system within the SMEFT framework will ultimately require the inclusion of full EW radiative corrections, which have been shown to induce the most significant distortions to the SM spin correlations in diboson processes~\cite{Grossi:2024jae,Goncalves:2025qem}. Such an extension goes beyond the current state-of-the-art and is left for future~work.

\section{Conclusions} \label{sec:conclusion}

In this article, we have presented new SMEFT predictions for diboson production and leptonic decays at the LHC. Specifically, we computed NLO QCD corrections for the full set of eight CP-even and -odd primary dimension-six operators that impact both gauge-boson self-interactions and Higgs-gauge-boson couplings. The necessary fixed-order SMEFT amplitudes were obtained using a modified version of the \recolatwo{} amplitude generator~\cite{Denner:2017vms,Denner:2017wsf}, which enables the selection of specific helicity configurations of the intermediate gauge bosons. This separation is achieved via a pole approximation, ensuring gauge-invariant results. The NLO QCD corrections were matched to~a~PS using the \POWHEGBOXRES{} method~\cite{Nason:2004rx,Frixione:2007vw,Alioli:2010xd,Jezo:2015aia}. Our new MC implementation therefore allows for a realistic and fully exclusive simulation of diboson production and their leptonic decays at NLO~QCD accuracy within the SM and beyond. 

To motivate simplified SMEFT benchmark scenarios, we reviewed the leading constraints on the Wilson coefficients of the dimension-six operators listed in~\cref{eq:operators}. In the case of the effective Higgs-gauge-boson interactions governed by $Q_{HB}$, $Q_{HW}$, $Q_{HW\!B}$, and their CP-odd counterparts, we found that the combination of the measured $W$-boson mass and the Higgs signal strength in $h \to \gamma \gamma$ imposes strong constraints on the associated Wilson coefficients. To evade these constraints, we have introduced the SMEFT benchmark scenarios~\cref{eq:CPevenbench,eq:CPoddbench}. The former predicts SM-like rates for $h \to WW$, $h \to ZZ$, and $h \to \gamma \gamma$, while enhancing the $h \to \gamma Z$ rate by approximately a factor of two compared to the SM. This pattern of deviations is favored by LHC~Run~2~data~\cite{ATLAS:2020qdt,CMS:2020gsy,ATLAS:2023yqk}. For~the two dimension-six operators involving only gauge-boson interactions, the constraints from $h \to \gamma \gamma$ are relatively weak, as the effects of $Q_{W}$ and $Q_{\widetilde{W}}$ only appear at the one-loop level. We~also pointed out that EDM measurements generally place very stringent constraints on all the CP-odd operators considered. However, these bounds can be relaxed in the SMEFT realizations where the Yukawa couplings of the light fermions vanish or are strongly suppressed. Given this model dependence, we have not considered the constraints on the CP-odd operators in~\cref{eq:operators} that could result from low-energy measurements in our phenomenological analysis of diboson production. 

We then conducted a comprehensive phenomenological study of the $pp \to W^+ Z \to e^+ \nu_e \hspace{0.5mm}\mu^+ \mu^-$ process to examine the influence of SMEFT contributions, based on the benchmark scenarios outlined earlier. As a first step, we validated the individual components of our new MC code by successfully reproducing the SMEFT results reported in~Refs.~\cite{Chiesa:2018lcs,ElFaham:2024uop}, and by employing a Legendre-projection technique to extract the polarization fractions of the intermediate gauge bosons. We~then carried out a thorough analysis of both singly- and doubly-polarized signals. For singly-polarized signals, we observed that the $Q_W$ and $Q_{\widetilde{W}}$ operators, relative to the SM, lead to a uniform shift~$\big($$\sin \hspace{0.25mm}(2\phi^\ast_{e^+})$ modulation$\big)$ in the longitudinal (transverse) component of the azimuthal decay angle distribution of the positron from the $W$ decay. For the benchmark choices $C_W/\Lambda^2 = 1 \, {\rm TeV}^{-2}$ and $C_{\widetilde{W}}/\Lambda^2 = 1 \, {\rm TeV}^{-2}$, the modulation amplitude reaches a relative magnitude of approximately~$3\%$. For the remaining operators, we find that they either do not contribute to $pp \to W^+ Z$ production at all, as is the case for $Q_{HB}$, $Q_{HW}$, and their CP-odd counterparts, or their contributions become negligibly small once other constraints, such as those from $h \to \gamma \gamma$, are taken into account, as~with~$Q_{HW\!B}$ and~$Q_{H\widetilde{W}\!B}$. In the case of doubly-polarized signals, we found that the operators~$Q_W$ and~$Q_{\widetilde{W}}$ do not contribute to the double-longitudinal polarization amplitudes in $W^+ Z$ production. The interference between the SM and the operator $Q_W$ reaches $3\%$ for the~$LT$~and~$TL$~polarization states, with opposite signs. The quadratic term boosts the $TT$ signal by $15\%$ via the SM-suppressed $W^+_\pm Z_\pm$ helicities. In~the mixed channels, linear and quadratic contributions are comparable. For~$Q_{\widetilde{W}}$, linear CP-odd effects are below $1\%$, while the quadratic terms rival or exceed the CP-even~ones. The $Q_W$ operator leads to a quadratic enhancement of the $TT$ contribution in the high-energy region of the $m_{3\ell\nu}^{\rm rec}$ distribution, while the $|\Delta y_{e^+Z}|$ observable reveals polarization-dependent BSM effects across both central and forward regions. These effects can significantly modify the polarization fractions, increasing the $TT$ component by up to $10\%$. We~have also shown that, in general, the impact of PS effects on SMEFT predictions closely mirrors that in the SM. A notable exception is the $p_{T,WZ}$ spectrum for the $TT$ signal, where the inclusion of quadratic contributions from the $Q_W$ operator leads to a significant reduction due to PS effects compared to the SM prediction.

Our MC implementation also enables, for the first time, a reliable quantum tomography of diboson systems in the presence of SMEFT effects, as well as higher-order corrections from QCD and PS matching. Specifically, we have computed fixed-order NLO QCD and NLO+PS results within the DPA for the most relevant spin-correlation coefficients involving one and two bosons. We observed that the coefficients $\alpha^{(V)}_{lm}$ and $\gamma_{lm l^\prime m^\prime}$ remain largely unchanged from their fixed-order values. A comprehensive analysis aimed at evaluating spin entanglement and Bell non-locality markers for qutrits is an important future direction, but lies beyond the scope of this work.

While the studies in this article are limited to $W^\pm Z$ production, we stress that our MC code also supports the simulation of $W^+ W^-$ production with leptonic decays, including the complete set of SMEFT operators listed in~\cref{eq:operators}. It could therefore be employed to perform precision polarization analyses of $W^+ W^-$ inclusive production at the LHC, both within the SM and in the presence of SMEFT~effects. Moreover, the results presented in this article provide valuable input for experimental efforts aimed at reducing modeling uncertainties in diboson polarization analyses at LHC~Run~3 and the HL-LHC. In particular, template-based, model-independent searches for new physics can significantly benefit from the new \POWHEGBOXRES{} implementation of SMEFT effects in diboson processes. Finally, extending the code to compute SMEFT contributions to diboson production at NNLO+PS, specifically for operators generating anomalous triple-gauge-boson couplings, is relatively straightforward using the techniques and results reported in~\citere{Gauld:2023gtb}. We defer such an extension to future~work.

\section*{Code availability}

The code used to perform the MC simulations in this paper is publicly available in a dedicated folder of the \POWHEGBOXRES{} GitLab repository:
\begin{center}
\url{https://gitlab.com/POWHEG-BOX/RES/User-Processes/VV_pol}
\end{center}

The relevant input parameters for simulating polarized-boson events within the SM and SMEFT frameworks, using the new \POWHEGBOXRES{} implementation, are listed in~\cref{app:MC}.

\section*{Acknowledgments}
The authors acknowledge support from the COMETA EU COST Action (CA22130). 
The work of GP and ER is partially supported by the Italian Ministero dell’Universit\` a e della Ricerca (MUR) through the research grant 20229KEFAM (PRIN2022, Finanziato dall’Unione europea - Next Generation EU, Missione 4, Componente 1, CUP H53D23000980006). GP acknowledges support from the EU Horizon Europe research and innovation programme under the Marie-Sk\l{}odowska Curie Action (MSCA) ``POEBLITA - POlarised Electroweak Bosons at the LHC with Improved Theoretical Accuracy'' - grant agreement Nr.~101149251 (CUP H45E2300129000). The MC simulations carried out in this article utilized computational resources provided by the Max Planck Computing and Data Facility (MPCDF). The Feynman diagrams shown in this work have been generated and drawn with \noun{FeynArts}~\cite{Hahn:2000kx}. 

\appendix

\section{Event generation with \POWHEGBOXRES{}}
\label{app:MC}

The phenomenological studies presented in this work rely on simulations that utilize the MC chain detailed in~\cref{sec:recola,sec:POWHEGBOXRES}. The usage of our new \POWHEGBOXRES{} implementation is perhaps best illustrated through a concrete example. The package includes a \verb|testrun| directory containing a sample \verb|powheg.input| file, which controls the event generation. 

The block in the sample \verb|powheg.input| file that defines the values of the Wilson coefficients has the following form:
\begin{verbatim}
 
 CHBD6 = 0.0
 CHWD6 = 0.0
 CHWBD6 = 0.0
 CWD6 = 0.0
 CHBtildeD6 = 0.0
 CHWtildeD6 = 0.0
 CHWtildeBD6 = 0.0
 CWtildeD6 = 1.0d-6

\end{verbatim}
The correspondence between the keywords and the Wilson coefficients is given by, for example, $\verb|CHBD6| = C_{HB}/\Lambda^2$, with analogous definitions for the remaining seven operators in \cref{eq:operators}. The~units of \verb|CHBD6| and the other keywords listed above are ${\rm GeV}^{-2}$. The above example therefore assigns $C_{\widetilde{W}}/\Lambda^2 = 1 \, {\rm TeV}^{-2}$, while all remaining Wilson coefficients are set to zero. 

Two other keywords related to the SMEFT part of the code are:
\begin{verbatim}

 NP_POWER = 1
 SUM_AMP = 1

\end{verbatim}
The keyword \verb|NP_POWER| specifies the power of $1/\Lambda^{2}$ up to which SMEFT contributions are included in the MC generation. Setting \verb|NP_POWER = 0| includes only the SM contribution, while setting \verb|NP_POWER = 1|, both the SM amplitude and its interference with the SMEFT amplitude are computed. Finally,~\verb|NP_POWER = 2| also calculates the squared SMEFT amplitude. The keyword \verb|SUM_AMP| determines whether the SM and SMEFT contributions --- up to the mass dimension specified by \verb|NP_POWER| --- are added together. Setting \verb|SUM_AMP = 1| enables the summation, while \verb|SUM_AMP = 0| only includes the contribution specified by the keyword \verb|NP_POWER|. In~the example above, the choice of keywords results in the computation of the sum of the SM amplitude and the SMEFT contribution linear in the Wilson coefficients, while the quadratic SMEFT contribution is excluded.

The selection of the specific helicity configurations for the intermediate gauge bosons is controlled by the following three keywords:
\begin{verbatim}

 dpa = 1
 pol1 = 1
 pol2 = 1

\end{verbatim}
The keyword \verb|dpa| can be set to 1 to activate the DPA, or to 0 to deactivate it. The keyword \verb|pol1| specifies the polarization of the first intermediate gauge boson. It can take the values $0$, $-1$, $+1$, $3$, or $4$, corresponding to longitudinal, left-, right-handed, transverse, or no polarization, respectively. \verb|pol2| performs the same function for the second intermediate gauge boson. The choices of the \verb|pol1| and \verb|pol2| keywords only influence the results when \verb|dpa = 1|. Otherwise, unpolarized predictions are generated. Except for the aforementioned, our new code can be used as described in the various documents within the \verb|Docs| directory of the standard \POWHEGBOXRES{} installation.


\providecommand{\href}[2]{#2}\begingroup\raggedright\endgroup

\end{document}